\documentclass[amsmath,amssymb,aps,superscriptaddress]{revtex4-1}
\usepackage{graphicx}
\usepackage{dcolumn}
\usepackage{bm}
\usepackage{overpic}

\usepackage{xcolor}

\newcommand{\ve}[1]{\ensuremath{\mbox{\boldmath$#1$}}}

\usepackage{amsmath,amssymb}

\begin{document}
\title{Smart Inertial Particles}
\author{Simona Colabrese}
\affiliation{Department of Physics and INFN, University of Rome ``Tor Vergata'', Via della Ricerca Scientifica 1, 00133, Rome, Italy.}
\email{simona.colabrese@roma2.infn.it}
\author{Kristian Gustavsson}
\affiliation{Department of Physics and INFN, University of Rome ``Tor Vergata'', Via della Ricerca Scientifica 1, 00133, Rome, Italy.}
\affiliation{Department of Physics, University of Gothenburg,  Origov\"agen 6 B, 41296, G\"oteborg, Sweden.}
\author{Antonio Celani}
\affiliation{Quantitative Life Sciences, The Abdus Salam International Centre for Theoretical Physics, Strada Costiera 11, 34151, Trieste, Italy.}
\author{Luca Biferale}
\affiliation{Department of Physics and INFN, University of Rome ``Tor Vergata'', Via della Ricerca Scientifica 1, 00133, Rome, Italy.}
\date{\today}
\begin{abstract}
  We performed a numerical study to train smart inertial particles to target specific flow regions with high vorticity
  through the use of reinforcement learning algorithms. The particles
  are able to actively change their size  to modify their inertia and density.
  In short, using local measurements of the flow vorticity, the smart particle explores the interplay between its choices of size and its dynamical behaviour in the flow environment. This allows it to accumulate experience and learn approximately optimal strategies of how to modulate its size in order to reach the target high-vorticity regions.
  We consider flows with different complexities:
  a two-dimensional stationary Taylor-Green like
  configuration, a two-dimensional time-dependent flow, and  finally
  a three-dimensional flow given by the stationary
  Arnold-Beltrami-Childress (ABC) helical flow. We show that smart particles are able to learn how to reach extremely intense vortical structures in all the tackled cases.\footnote{\textbf{Postprint version of the article published on Phys. Rev. Fluids 3, 084301 (2018) DOI: https://doi.org/10.1103/PhysRevFluids.3.084301}}
\end{abstract}
\maketitle
\section{Introduction}
Controlling and predicting the dynamical  evolution and spatial distribution of small particles suspended in complex flows is
a fundamental  problem in many  applied disciplines such as in spray combustion,  drug delivery,  dispersion of pollutants or contaminants in the environment, spray formation and rain formation in clouds, to cite just a few cases \cite{mostafa1987modeling,faeth1996spray,patrick2012modeling,kovetz1969effect,langer1990new,boekerdispersion}. The dynamics of small particles is also
used in turbulence to study the Lagrangian statistics of the flow, and/or the instantaneous Eulerian velocity distribution in Particles Image Velocimetry techniques \cite{toschi2009lagrangian,adrian2011particle}. Small particles have been
instrumented to perform local measurements of flow properties \cite{shew2007instrumented}. In this paper we present a numerical study to show
how one can implement a suitable learning algorithm to train micro particles to actively control their dynamical trajectories in order to  achieve some predetermined goal, for example reaching a very intense vorticity region in the flow, escaping from turbulent fluctuations, preferentially tracking specific topological structures etc.
The number of potential applications is high, from engineering of small particles that are able to perform simple tasks by measurements of specific turbulent fluctuations, to small active engines able to modify, actuate, react on the flow structure both at micro- and macro-scales \cite{ma2016reversed,baraban2012catalytic,solovev2011light,walker2006micro,wynn2014autonomous,calzavarini2018propelled}. On a macroscopic level, there are key applications in marine and atmospheric research, where buoys and small balloons are instrumented to measure local field properties~\cite{basso2017disposable,argo,weather}. It is crucial to optimise their trajectories during the diurnal cycles, but this is often a complicated task due to large fluctuations in the surrounding flow.
Another application lies in understanding, predicting, and taking advantage of the behaviour of biological organisms~\cite{muinos2018reinforcement}. One example being diatom algae that are known to be able to adjust their densities in the stratified turbulent ocean in order to perform vertical migration \cite{sarthou2005diatoms,gemmell2016buoyancy,walsby1994gas,arrieta2015microscale}.


It is well known that particles that are heavier or lighter than the fluid
systematically detach from the flow
streamlines~\cite{maxey1983equation,bec2005multifractal,balkovsky2001intermittent,bec2010turbulent,biferale2005particle}. 
As a consequence, correlations between particle positions and structures of the underlying flow appear.
Heavy particles are expelled from vortical structures, while light particles tend to concentrate in their cores.
This results in the formation of strong inhomogeneities in the spatial distribution of the particles, an effect often referred to as preferential concentration~\cite{douady1991direct,squires1991preferential,eaton1994preferential,monchaux2012analyzing,bec2007heavy,boffetta2004large}.

Thanks to these properties,
light particles have been used as small probes
that preferentially track any high-vorticity structure, highlighting statistical and topological properties of the underlying fluid conditioned on those structures \cite{calzavarini2006microbubble,milenkovic2007bubble}.

In this paper, we seek inertial particles capable of
sampling \textit{ad-hoc} specific flow  structures.
We imagine smart particles to be  endowed with the ability of obtaining some partial information about the regions of the flow that they
are visiting. They can use this knowledge to learn how to adapt their size and consequently their inertia and density in order to preferentially
sample only some predetermined flow properties. 

The set of questions we  want to address are the following: can these smart particles learn how to track specific targets in an approximately optimal way in complex flows? Is this achievable without any previous knowledge of the flow structures and by using a set of simple behavioral actions the particle may take? Is learning achievable  even when
tracer particles would evolve in a chaotic and unpredictable way? Is it possible to guess the typical form of the approximately optimal strategies a priori? To what extent are the approximately optimal strategies learnt in a stationary flow environment also robust by adding time-dependence to the flow?

Similar questions have been investigated by training based on reinforcement learning algorithms for the
dynamics of smart micro-swimmers in Taylor Green flows \cite{colabrese2017flow}, in Arnold-Beltrami-Childress (ABC) flows \cite{gustavsson2017finding} and  for the case of fish schooling both in still water and vortical flows \cite{gazzola2016learning,gazzola2014reinforcement,novati2017synchronisation,verma2018efficient}. A similar approach has also been pioneered in \cite{reddy2016learning} for the case of  birds that  can exploit warm thermals to soar in a turbulent environment.
Reinforcement learning is a framework to find good strategies for achieving given long-term tasks. It is widely used in artificial intelligence and machine learning. It is based on the interaction between a decision-maker (in our case the inertial particle) and the environment. The decision maker can change its behaviour in response to inputs from the system. By trial and error the decision maker progressively learns how to behave optimally \cite{sutton2017reinforcement,tesauro1995temporal,kaelbling1996reinforcement}.
We show that this approach provides  a way to construct efficient strategies by accumulating experience also for the cases investigated here.

The paper is organized as follows. In Section \ref{sec:in} we define the model used to describe inertial particles. It will be the groundwork for the designing of smart  particles. In Section \ref{sec:RL} we provide an overview on the reinforcement learning approach and on the algorithm used. In Section \ref{sec:TGflow} we discuss  the application of the reinforcement algorithm to particles in a Taylor Green-like flow. It has given space to the algorithm implementation and to the results achieved, both for stationary and time-dependent flows. Similarly, Section \ref{sec:ABC} deals with the case of ABC flows.
Finally, in Section \ref{sec:co} we draw the conclusions and final remarks of the paper.

\section{Inertial Particles}
\label{sec:in}
We focus the discussion to the case of small spherical inertial particles that have a density that is different from the surrounding fluid.
We consider only the diluted limit, neglecting possible
hydrodynamical interactions, collisions and aggregation among the particles. The particles  are small enough to be
considered as point-like. They are characterized by their mass, $m_{\rm p}$ and by their {\it adjustable} radius $b$.
A commonly used model for inertial particles with arbitrary density is the following~\cite{auton1988force,babiano2000dynamics,gatignol1983faxen,maxey1983equation}
\begin{equation}\label{eq:model}
  \begin{cases}
    \dot{\textbf{X}} =  \textbf{V} + \sqrt{2\chi}\ve\eta, \;\\
    \dot{\textbf{V}} =  - \frac{1}{St}(\textbf{V} - \textbf{u}(\textbf{X},t)) + \beta D_t \textbf{u}(\textbf{X},t)\,.
    \end{cases}
\end{equation}
Here dots denote time derivatives, $ \textbf{X} $ and $ \textbf{V} $ are dimensionless particle position and velocity,
$ \textbf{u}$ and $D_t\textbf{u} = \partial_t \textbf{u} + (\textbf{u}\cdot \nabla)\textbf{u} $ denote dimensionless flow velocity field and material  derivative evaluated at the particle position,  $\ve\eta(t)$ is a Gaussian white noise $\langle \eta_i(t)\eta_j(t') \rangle = \delta_{ij}\delta(t-t')$ with a small prefactor $\chi$ added to avoid structurally unstable solutions in the two-dimensional steady flow.\\
The dimensionless Stokes number
\begin{equation}\label{eq:stokes}
{\rm St} = \tau_{\rm p} / \tau
\end{equation}
in (\ref{eq:model}) is defined in terms of the ratio of the characteristic flow time $\tau$ and the  particle response time:
\begin{equation}\label{eq:tau}
\tau_{\rm p} = b^2/(3 \nu \beta)
\end{equation}
where  $\nu$ is the kinematic viscosity  of the carrying flow. The dimensionless quantity
\begin{equation}\label{eq:beta}
\beta = 3\rho_{\rm f}/(\rho_{\rm f} + 2\rho_{\rm p})
\end{equation}
accounts for the added mass effect resulting from the contrast between the particle density,
$ \rho_{\rm p} = 3 m_{\rm p}/ (4 \pi b^3)$, and the fluid density $ \rho_{\rm f}$.
The value $\beta=1$ distinguishes particles that are heavier ($\beta<1$) or lighter ($\beta>1$) than the fluid.
The dimensionless translational diffusivity $\chi$ in (\ref{eq:model}) is taken to be small, $\chi\ll 1$.
For Eqs.~(\ref{eq:model}) to be valid, it is assumed that the particle size $b$ is much smaller than the smallest
active scale of the flow, and that the  Reynolds number based on the particle size, ${\rm Re}_{\rm p}\equiv |\textbf{u}-\textbf{V}| b/\nu$,
is very small, ${\rm Re}_{\rm p} \ll 1$.
In the limits ${\rm St}\to 0$ or $\beta\to 1$ the dynamics determined by Eqs.~(\ref{eq:model}) tends to the evolution of a tracer: imposing a finite Stokes drag leads to $ \textbf{V} = \textbf{u} + O({\rm St}(1-\beta))+O(\chi)$.
\\
Both St and $ \beta $ depend on the radius $ b $. As a consequence, particles that are able to modify their own radius depending on cues from the surrounding environment will indirectly exercise a control over their trajectory and may be able to alter their dynamics to navigate the flow. We consider a liquid-solid system in which particles can adjust their size $ b $ by choosing among a set of discrete values to be able to be either lighter or heavier than the fluid (see Table \ref{tab:stparameter} and Subsec. \ref{sub:Ai}). 
The equation system (\ref{eq:model}) is discretized using the Runge-Kutta 4th order method with $ dt= 0.001 $ (much smaller than all the other time scales in the system). 
\section{Reinforcement Learning Approach}
\label{sec:RL}
\begin{figure*}[htbp]
	\begin{overpic}[scale=0.35,trim={0mm 160mm 0mm 0mm}]{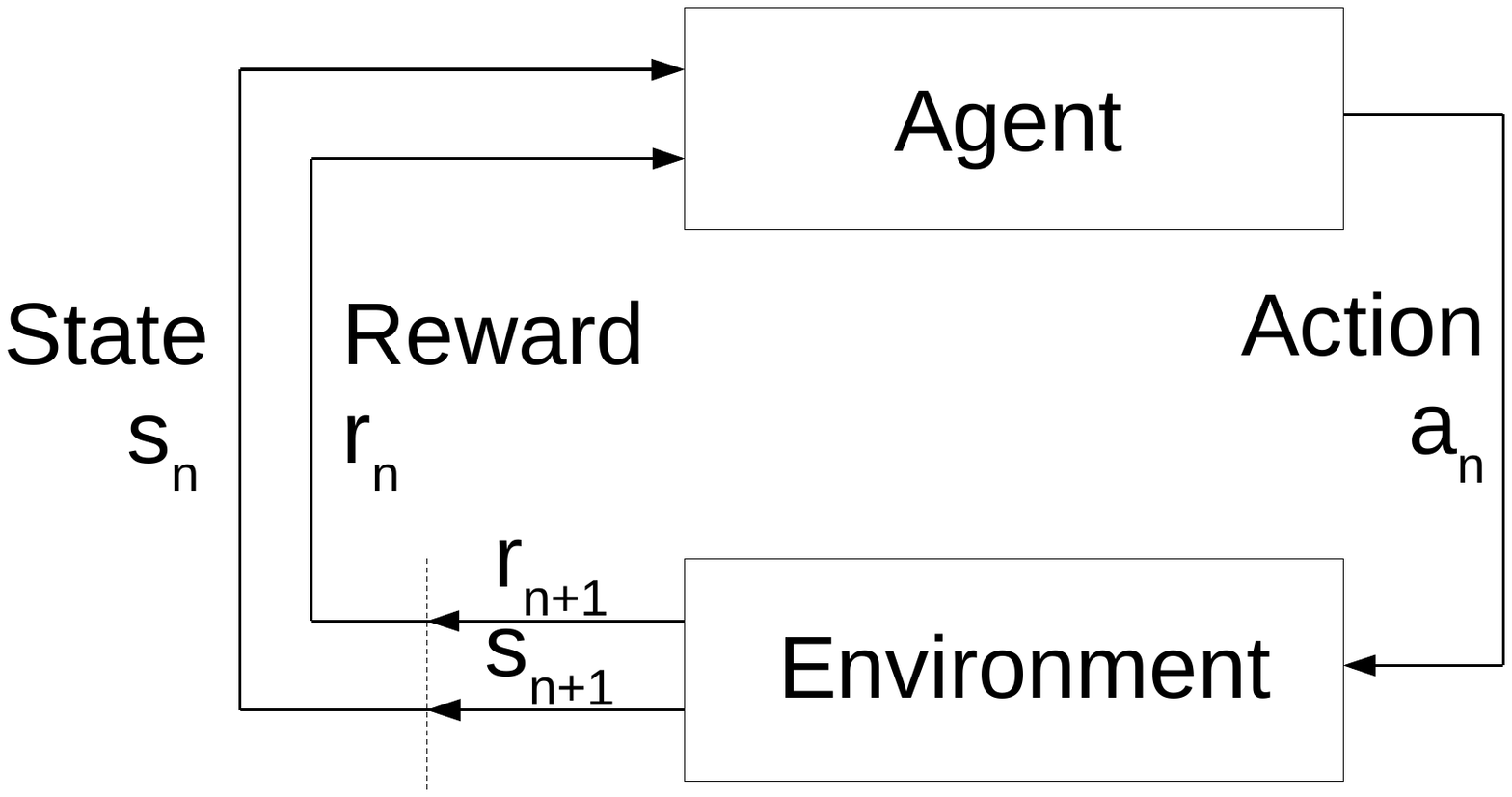}
    \put(0,47){${\bf a}$}
    \end{overpic}
	\begin{overpic}[scale=0.5,trim={0mm 0mm 0mm 0mm}]{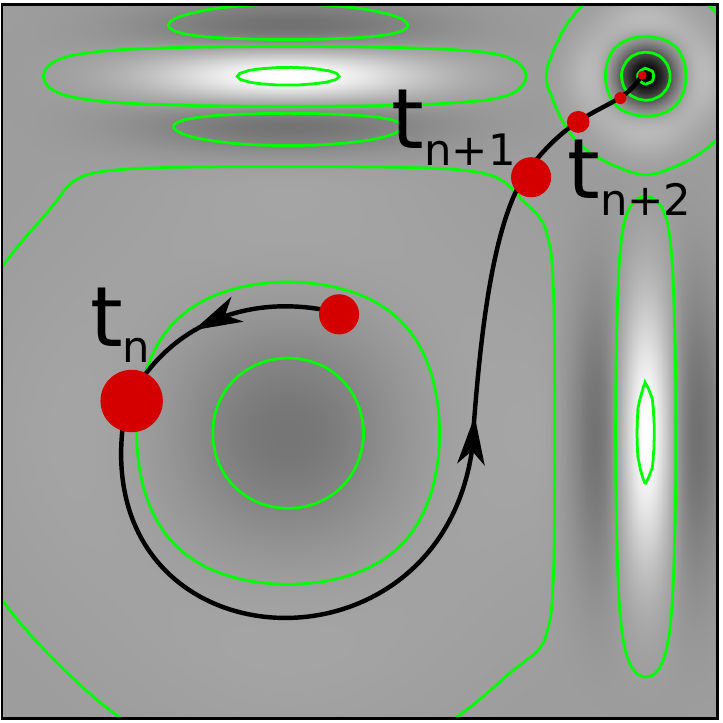}
    \put(-10,92){${\bf b}$}
    \end{overpic}
	\caption{{\bf a} Sketch of the reinforcement learning agent-environment interactions. {\bf b}
          Graphical summary of a typical trajectory for a smart particle in a generic two-dimensional flow.
          State changes occur at times $t_n$, $t_{n+1}$, $t_{n+2}$, \dots when the particle enters new vorticity regions that are separated by isolines of the vorticity field $\Omega_z$ (green lines).
          In each state the particle chooses its size (the action).
          }\label{fig:diagram}
\end{figure*}

To identify efficient strategies to sample high-vorticity regions in complex flows, we used reinforcement learning to implement the {\it one-step $Q$-learning} algorithm~\cite{watkins1992q}.
The general reinforcement-learning framework consists of an {\it agent} that is able to interact with its environment (see Fig.~\ref{fig:diagram}{\bf a} for a schematic illustration).
At any given time, the agent has the ability to sense some information about the environment, or about itself. This information forms the {\it state} $s$, which is an element of the set ${\mathcal S}$ consisting of all the possible distinct states the agent can recognize.
Depending on the current state $s_n$, the agent chooses an {\it action} $a_n$ from a set ${\mathcal A}$ of possible actions.
Which action is chosen affects the interaction between the agent and the environment.

 The transition $ s_n \rightarrow s_{n+1} $ occurs either if the agent encounters a new state ($ s_{n+1} \neq s_n $) or if a maximum time, $ T_{max} $, has run out (in the latter case $ s_{n+1} = s_n $). 
The time $T_{max}$ is chosen to be much larger than the characteristic flow time scale, $T_{max} \gg \tau$.
	After the transition the agent is given a {\it reward} $r_{n+1}$ 
quantifying the immediate success of the previously chosen action to reach the target goal.

It is important to remark that in our way of implementing the reinforcement learning algorithm the physical time between consecutive state changes may fluctuate depending on the underlying dynamics. In principle, this implies the possibility for the modeled particle to have a nearly continuous-in-time control on its state if state changes occur frequently enough. An alternative implementation is based on a fixed time interval between two consecutive probes of the ambient state.The latter protocol is suitable for algorithms based on a continuous representation of the state-action space. This protocol is also required to ensure the possibility of reaching one of the optimal policies in the case in which the process is Markovian. In our case, as in most realistic applications, the process is not Markovian. States are vorticity levels and not positions in space; therefore, there is no unique recipe or best practice for modeling transitions. For the sake of this work, we only explored the first protocol explained above, without conditioning state changes on a fixed time interval. We checked \textit{a posteriori} that for the quasi optimal policy the most probable time interval between two consecutive state changes coincides with the upper limit $ T_{\mbox{max}} $ ($ 70 \% $ of all cases), so very frequent controls are not the typical ones.
 
Depending on the reward, the agent updates its {\it policy} of how to select the action for any given state.
Finally, the agent chooses an action $a_{n+1}$ for the new state $s_{n+1}$ using the updated policy and repeats the procedure outlined above (see Fig.~\ref{fig:diagram}{\bf a}).
The aim of reinforcement learning is to form a close-to optimal policy of how to choose the action for a given state to achieve the predetermined goal. This is done by maximizing the long-term accumulated reward.

In our case, the agent is the smart inertial particle.
It has as a target to navigate vortex flows to reach regions of high vorticity.
One example trajectory in a generic two-dimensional vortex flow is sketched in Fig.~\ref{fig:diagram}{\bf b}.
We assume that the particle can sense the $z$-component $\Omega_z$ of the local flow vorticity $\ve\Omega=\ve\nabla\times\ve u$.
It can distinguish a discrete number $N_s$ of coarse grained states corresponding to equally spaced intervals of $\Omega_z$ in the full range of $\Omega_z$ allowed by the flow.
Different states are separated by isolines of the flow vorticity as illustrated in Fig.~\ref{fig:diagram}{\bf b}.
Each time $t_n$ the particle enters a new state $s_n$ it selects a size (the action $a_n$) according to the current policy out of a finite discrete set of $N_a$ possible sizes:
$$a_n \in {\cal A}=\{ b\}_1^{N_a}\,.$$
As a result, the radius of the particle is a dynamical variable, $b \to b_n$, leading to a change in the particle density $\beta \to \beta(b_n)$ and inertia ${\rm St} \to {\rm St}(b_n)$.
Depending on which sizes the particle chooses, it will in general experience a different dynamical evolution.
The particle keeps its size until the time $t_{n+1}$ where it enters the next state $s_{n+1}$ and is given a reward $r_{n+1}$.
In order to train the particle to move into regions of high vorticity, we choose the reward to be proportional to some power of the vorticity.

The core of the learning protocol lies in the policy $\pi$ of how to chose an action given a state, $\pi(s): s  \to a$.
To find an approximately optimal policy a {\it training phase} is performed.
During this phase, the particle is allowed to explore the effects of taking different actions in the different states.
We use the one-step $Q$-learning algorithm to find an approximately optimal policy, $\pi^*$, iteratively by introduction of intermediate policies $\pi_n$ after the $n$-th state change, such that $\lim_{n\to\infty}\pi_n=\pi^*$.
The policy $\pi_n$ is derived from the {\it $Q$-value}, $Q_{\pi}(s_n,a_n)$ (hence the name of the algorithm), according to the {\it $\epsilon$-greedy} rule:
select the action that maximizes the current $Q$-value function except for a small probability $\epsilon$ of randomly selecting another
action independent of the $Q$-value:
\begin{align}\label{eq:1}
	a_{n+1} =\bigg \{
	\begin{array}{ll}
		{\rm arg}\max_{a'}Q(s_{n+1}, a') &{\rm with \quad probability \quad } 1-\epsilon \\
		{\rm a\quad random \quad action} & {\rm with\quad probability\quad} \epsilon
	\end{array}\,.
\end{align}
For every state-action pair $(s_n,a_n)$ at time $t_n$, the $Q$-value estimates the expected sum of future rewards conditioned on the current status of the system and on the current policy $\pi_n$:
\begin{equation}
\label{eq:return}
Q_{\pi_n}(s_n,a_n) = \left\langle r_{n+1} + \gamma r_{n+2} +
\gamma^2 r_{n+3} + \cdots\right\rangle\,.
\end{equation}
The parameter $ \gamma $ in Eq.~(\ref{eq:return}) is the {\it discount factor}, $ 0 \leq \gamma < 1$.
The value of $\gamma$ changes the resulting strategy:  with a myopic evaluation ($ \gamma $ close to $0$), the approximately optimal policy greedily maximizes only the immediate reward. As $ \gamma$ gets closer to $1$, later rewards contribute significantly (far-sighted evaluation).

Each time a state change occurs, the agent is given a reward $r_{n+1}$, which is used together with the value of the next state $s_{n+1}$ to update the $Q(s_n,a_n)$-value according to the following rule \cite{sutton2017reinforcement}:
\begin{equation}\label{eq:Q}
	Q(s_n, a_n) \leftarrow  Q(s_n, a_n) +
	  \alpha [r_{n+1} + \gamma \max_{a}Q(s_{n+1}, a) - Q(s_n, a_n)]\,,
\end{equation}
where $ \alpha $ is a parameter that tunes the learning rate. The learning estimate is based on the differences between temporally successive predictions. In particular, in one-step $Q$-learning (\ref{eq:Q}), between the current prediction, $Q(s_n, a_n)$, and the one that follows, $r_{n+1} + \gamma \max_{a}Q(s_{n+1}, a)$. For Markovian systems and if $\epsilon$ slowly approaches zero as $n\to\infty$, it is possible to show that the update rule (\ref{eq:Q}) converges to an optimal $Q(s,a)$ with a derived policy $\pi^*(s)$ which assign to each state the action maximizing the expected long term accumulated reward~\cite{sutton2017reinforcement}.
  The system of inertial particles considered here is not Markovian, but still we expect approximately optimal policies to be found by the one-step $Q$-learning algorithm.

Operationally, we broke the training phases into a number $N_{\rm E}$ subsequent episodes $E$, with $E=1,\dots,N_E$.
The first episode is initialized with an
optimistic $Q$-value, i.e. all entries are equal and very large compared to the maximal achievable reward.
This has the effect to encourage exploration and to avoid to be trapped around local minima in the search for approximately optimal policies. Each episode ends after a fixed number of total state-changes, $N$, and it is followed by a new episode with a new random initial position of the particle in order to introduce further exploration. The initial velocity of the particle is equal to that of the fluid at the particle position. In order to accumulate experience,
the initial $Q$ of each new episode is given by the one obtained at the end of the previous episode.
For the purpose to quantify the learning ability of the smart particle during the training process, we monitor the total amount of reward that the particle gains in each episode:  $$\Sigma(E) = \sum_{n=1}^{N}r_n.$$
Finally, after the training has been performed, for each state $s$ the final policy is to choose the action $a$ with the maximal entry in the $Q$-value function.
To quantify the success of smart particles we use this policy to evaluate the long-term accumulative discounted reward, the {\it return}:
\begin{equation}
  \label{eq:reward}
R_{\rm tot} = \Big\langle \sum_{n=1}^{N} r_n \gamma^n\Big\rangle\,.
\end{equation}
Here the sum extends up to the total number of state changes, $N$, and the average is taken over realizations of the noise and over the initial conditions of Eqs.~(\ref{eq:model}).
The return $R_{\rm tot}$ can also be used during training to evaluate the success of the intermediate policies $\pi_n$.

\section{Application to a Taylor Green-like flow}
\label{sec:TGflow}
As a first example we studied smart inertial particles in a two-dimensional stationary flow.
 The flow is differentiable and incompressible everywhere in the considered domain,
 it consists of four main vortices of different intensity and sign, singled out by appropriate Gaussian functions (see Fig.~\ref{fig:vorticity}{\bf a} and Appendix I for a detailed description). This is a simple set-up to test the ability of the smart particle to discriminate between vortices of the same sign but with different basin of attraction and intensity.

\subsection{Algorithm implementation}
\begin{figure*}[htbp]
	\centering
	\begin{overpic}[scale=1,trim={20mm 24mm 0mm 5mm}]{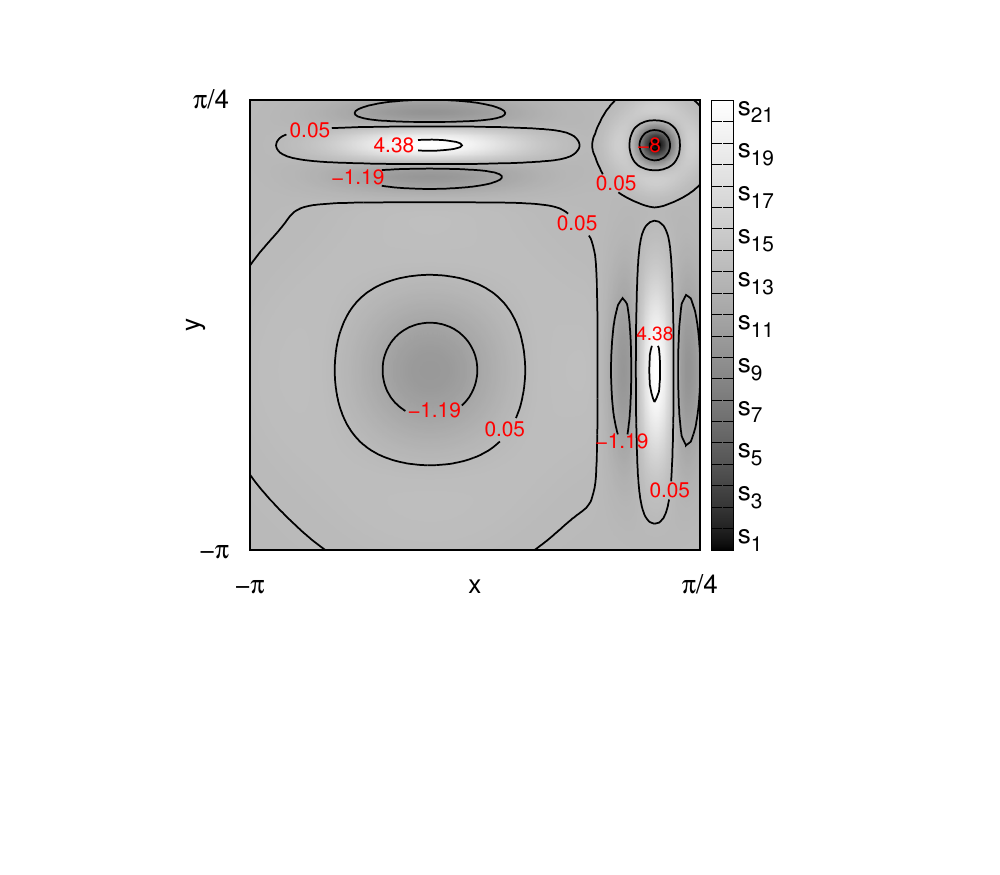}
    \put(-7,63){${\bf a}$}
    \end{overpic}
	\begin{overpic}[scale=1,trim={20mm 24mm 0mm 5mm}]{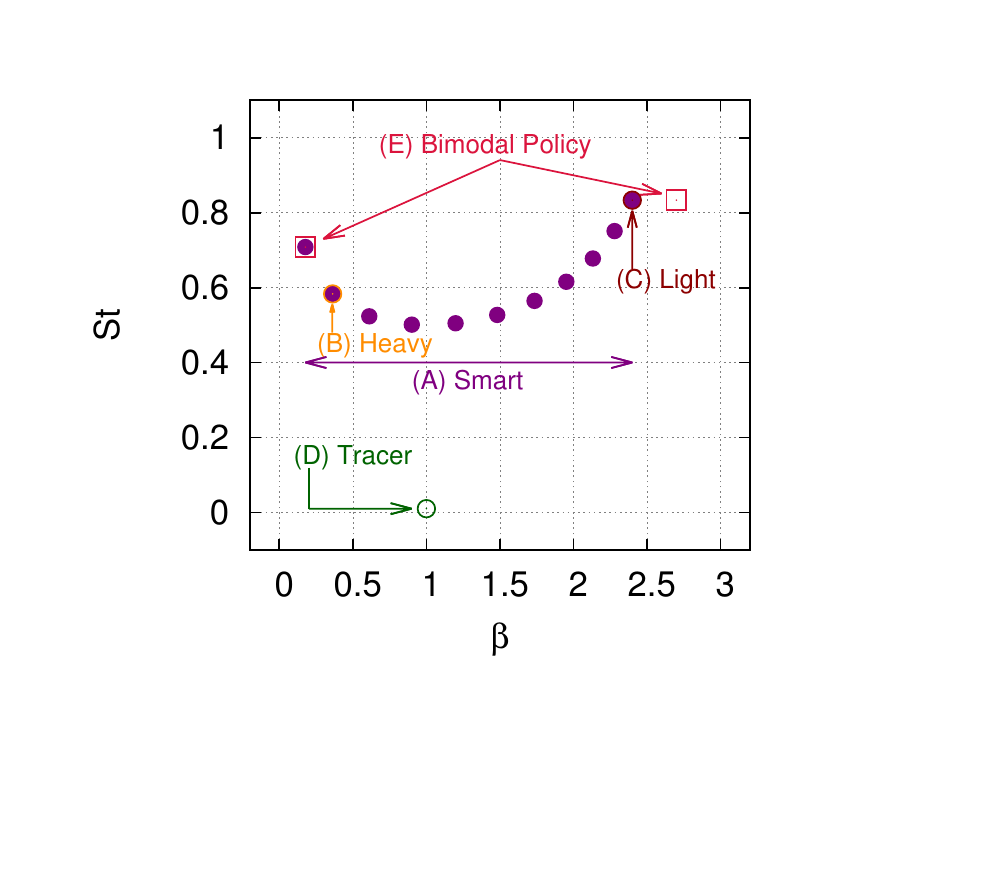}
    \put(-7,63){${\bf b}$}
    \end{overpic}
	\caption{ {\bf a} An illustrative sample of the isolines delimiting the $N_s=21$ vorticity states. Values corresponding to the underlying vorticity isolines are shown in red. 
	{\bf b} Set of pairs (${\rm St}$,$ \beta$) corresponding to the selection of the $N_a=11$ actions. Smart particles are allowed to select each one of the available actions freely (A). We also highlight four different cases corresponding to four possible naive strategies: (B)-Heavy particle with $({\rm St}=0.58,\beta=0.36)$ ;
	(C)-Light particle with $({\rm St}=0.83, \beta=2.4)$;
	(D)-Tracer with  $({\rm St}=0.01,\beta=1)$; (E) simple bimodal policy with only two actions:  the particle has
        one constant low density $({\rm St}=0.83,\beta=2.7)$ [light]
        in strong negative vorticity states or one constant high density $(St=0.71,\beta =0.18)$ [heavy] in all the other states (see also text).
        }\label{fig:vorticity}
\end{figure*}

We divide the scalar vorticity field into $N_s=21$ equally spaced states as sketched in Fig.~\ref{fig:vorticity}{\bf a}.
The only parameters affecting the motion in Eqs. (\ref{eq:model}) are $\beta$ and ${\rm St}$, given by Eqs.~(\ref{eq:stokes}-\ref{eq:beta}).
We assume that the smart particle can increase its size $b$ by a factor four compared to a smallest size, $b_{\rm min}$.
We use $N_a=11$ actions of equally distributed magnification factors compared to $b_{\rm min}$.
The values of $\beta$ and ${\rm St}$ corresponding to these actions are illustrated in Fig.~\ref{fig:vorticity}{\bf b} (fixing the values of $\beta$ and ${\rm St}$ at $b=b_{\rm min}$ such that the particle is heavy and with ${\rm St}$ of order unity). 
In Table~\ref{tab:stparameter} we report all values for $\beta$ and ${\rm St}$ for each one of the $N_a=11$
different sizes. The window of ${\rm St}$ is chosen in order to optimize the sensitivity to $ \beta $ of the particle trajectories \cite{bec2005multifractal,bec2007heavy}.
 \begin{table}[h]
 	\centering
 	\begin{tabular}{c | l l l | c}
 		&$b/b_{\rm min}$ &	$\beta$ & ${\rm St}$ \\
 		\hline
 		$a_1$&1 & 0.176 & 0.708 & Heavy\\
 		$a_2$&1.3 & 0.362 & 0.583 & \quad \quad Heavy (B)\\
 		$a_3$&1.6 & 0.611 & 0.523 & Heavy\\
 		$a_4$&1.9 & 0.900 & 0.501 & Heavy\\
 		$a_5$&2.2 & 1.20 & 0.505 & Light\\
 		$a_6$&2.5 & 1.48 & 0.527 & Light\\
 		$a_7$&2.8 & 1.74 & 0.565 & Light\\
 		$a_8$&3.1 & 1.95 & 0.615 & Light\\
 		$a_9$&3.4 & 2.13 & 0.678 & Light\\
 		$a_{10}$&3.7 & 2.28 & 0.751 & Light\\
 		$a_{11}$&4 & 2.40 & 0.833 & \quad \quad Light (C)
 	\end{tabular}
 \caption{Set of parameters corresponding to the $N_a=11$ actions. The second and the last rows of parameters indicate also the action values for the naive strategies. The dynamics in Eqs. (\ref{eq:model}) is affected only by the adimensional parameters $St$ and $\beta$. 
 All values of ${\rm St}$ and $\beta$ in the table are determined by fixing $\beta=0.176$ and ${\rm St}=0.708$ at $b=b_{\rm min}$.
 }\label{tab:stparameter}
 \end{table}
 
We contrast the smart particle to naive particles with simple strategies whose actions are illustrated in Fig.~\ref{fig:vorticity}{\bf b}: being a heavy particle (B), being a light particle (C), being a tracer particle (D), and being a bimodal particle that is light in flow regions of large negative vorticity and heavy otherwise (E).

To probe the efficiency of the algorithm, we choose the difficult task for the particles to reach the small upper right flow region in Fig.~\ref{fig:vorticity}{\bf a} independently of the initial condition and within a limited number of state changes, $N$ (here $N=5000$).
The upper right region has the highest negative vorticity of the flow. We therefore assign a reward proportional to  the cube of the negative vorticity  experienced by the particle when it crosses the border between two vorticity levels:
\begin{equation}
r_{n+1} = -s_{n+1}^3,
\end{equation}
where the minus sign is used to target negative vorticity regions.
Due to the presence of different peaks in the vorticity distribution the task is non-trivial. For example,  naive
light particles with a given fixed density and not too high Stokes number (for example case (C) in Fig.~\ref{fig:vorticity}{\bf b}) would simply be attracted by the vortex closest to their initial condition, independently of the sign and intensity of the vortex.

We consider reinforcement learning using the scheme described in Section~\ref{sec:RL} with  mainly fixed learning rate $\alpha = 0.1 $ and $\epsilon=0$ (for a case with time-dependent $\alpha$ and $\epsilon$ see Section \ref{sssec:exploration}). The other fixed parameters are $\gamma = 0.999$ and $\chi=0.005$. The parameters are empirically chosen to achieve the convergence to approximately optimal policies.
To enhance exploration, the elements of the initial Q-value matrix are chosen to be equal to the undiscounted return that a particle would gain if it was in the target region during the entire length of the episode: $Q_{\pi_0}(s,a) = -\Omega^3_{\rm min}N$, for all $(s,a)$.
 \subsection{Results}
\begin{figure}[htbp]
	\centering
	\includegraphics[scale=.8]{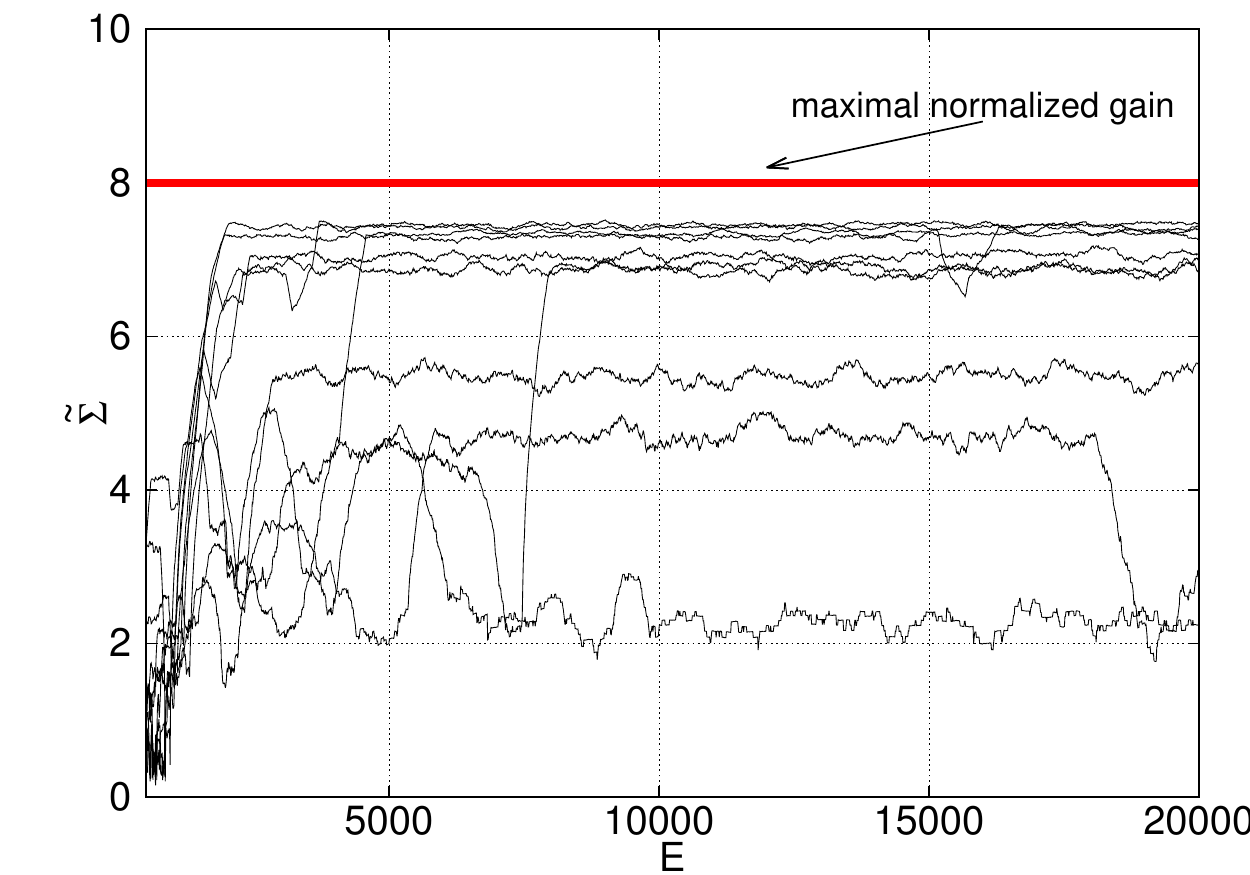}
	\caption{Dependence of the normalized total gain, $\tilde{\Sigma}$ in Eq.~(\ref{eq:normed_tot_return}) on the episode $E$
          for ten different learning processes (black curves). Every point represents an average over a sliding window of $500$ episodes. The red line shows the maximal possible reward.}\label{fig:Return}
\end{figure}

We first discuss the training session.
In Fig.~\ref{fig:Return} we show the evolution of the normalized total gain:
\begin{equation}
\tilde{\Sigma}(E) = \sqrt[3]{\frac{\Sigma(E)}{N}},
\label{eq:normed_tot_return}
\end{equation}
where the normalization is introduced such that the maximum achievable gain corresponds to the cube-root of the maximal reward, or equivalently to the minimal negative vorticity of the flow, $\Omega_{\rm min}=-8$, found in the upper right region in Fig.~\ref{fig:vorticity}{\bf a} (state $s_1$).
For each session, the smart particle increases its performance during the training phase and eventually achieves to reach the smallest vortex for most initial conditions and realizations of the white noise $\ve\eta$. Figure~\ref{fig:Return} shows the evolution of the learning gain as a function of the episode during the training process for $ 10 $ different trials.
We observe that the different trials result in different values of the normalized gain after many episodes.
The greedy choice ($\epsilon=0$) of actions based on $Q$ that we have adopted in this particular numerical experiment allow for little exploration.
After an initial transient where much exploration occurs, the evolution of the initially large $Q$-value matrix almost stagnates.
As a result it might happen that the dynamical relaxation (\ref{eq:Q}) toward the approximately optimal policy gets stuck for many training episodes in a local optimum.
The only way to leave the local optimum is given by the relatively small exploration due to the choice of initial condition and the realization of the noise $\ve\eta$.

\begin{figure}[htbp]
	\centering
	\includegraphics[scale=0.71,trim={110mm 20mm 100mm 0mm}]{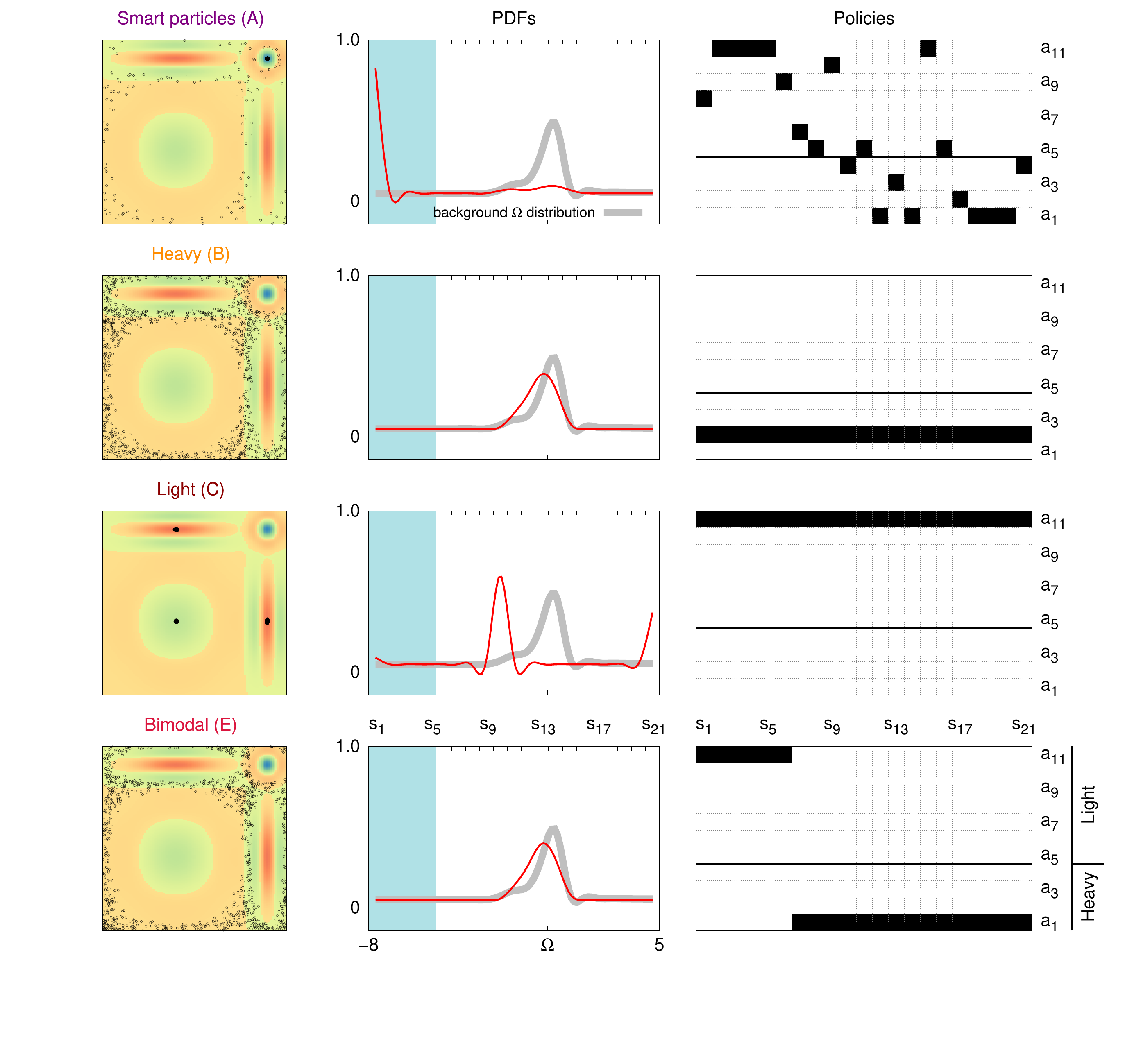}
	\caption{
          Left column: Particle positions along $10$ representative trajectories plotted after an initial transient for the cases studied in  Fig.~\ref{fig:vorticity}{\bf b} (except for tracer particles which fill space uniformly).
          From top to bottom: Smart-particles (A), Heavy (B), Light (C), simple bimodal policy (E).
          Middle column: probability density function of vorticity sampled along $1000$ particle trajectories for each of the cases studied (red curves). The PDFs are compared with the distribution for tracer particles (shaded curve).
          The blue shaded region marks vorticity levels that are unique to the target region. 
          Right column: the policies (the action $ a $ the particle takes given a state $ s $) used in the considered cases. 
          }\label{fig:Trajpdfqmatrix}
\end{figure}

After the training, we perform an exam session, by taking the policy derived from the final $Q$ obtained from one of the successful trials shown in Fig.~\ref{fig:Return}.
In Fig.~\ref{fig:Trajpdfqmatrix} we show the spatial distribution of smart particles, using the derived policy which gives the highest gain in Fig. \ref{fig:Return}.
This is compared to the spatial distribution for the four naive reference cases discussed above and shown in Fig. \ref{fig:vorticity}{\bf b}.
We observe that the trajectories of smart particles have high density in the target region, which isn't sampled at all in the other instances, or just rarely for the case of light particles. Next to the position plots, we show the probability density functions of the vorticity sampled by the particles (middle column),
which give a quantitative representation of the frequency at which the particle visits different states.
Notice how the smart particles (first row) are able to oversample the target of the training, the intense negative vorticity region, avoiding for most of their time the other lower-reward vortices.  It is important to remark that the approximately optimal policy obtained via the reinforcement learning is much better than the one with a bimodal change in density: being heavy for vorticities outside of the target region and light otherwise (fourth row in Fig.~\ref{fig:Trajpdfqmatrix}). We have also tested that the policy obtained by using the RL algorithm is always much better than any other bimodal policy defined with other states when the particles switch from light to heavy (not shown). In other words, to reach the small target it is necessary to take unintuitive actions even in a relatively
simple flow as the one considered here. This is summarized by the complex structure of the
approximately optimal policy shown in the right column of Fig.~\ref{fig:Trajpdfqmatrix}.
In Table~\ref{tab:Rcomparison} we report a quantitative
comparison between the long-term normalized return
\begin{equation}
\tilde{R}_{\rm tot} = \sqrt[3]{\frac{\gamma-1}{\gamma^{N+1}-1}}\sqrt[3]{R_{\rm tot}}\label{eq:norm_disc_ret}
\end{equation}
for smart particles and the other reference cases. The normalization used for $ R_{\rm tot} $ in (\ref{eq:norm_disc_ret}) is given by the sum of the first $ N $ terms of the geometric series with common ratio $ \gamma $.
\begin{table}
\centering
\begin{tabular}{l l| l l l l l}
	&  & (A)   & (B)  & (C)  & (D)  & (E) \\
	S-TG & $\tilde{R}_{\rm tot}$ & 7.49 & 0.76 & 2.4 & 1.4 & 1.6\\
	T-TG & $\tilde{R}_{\rm tot}$ & 5.0 & 0.0  & -2.0 & 0.0  & 0.0
\end{tabular}\caption{ The normalized discounted return, $\tilde{R}_{\rm tot}$ for the best training of the smart particles using a greedy policy, compared with the other four cases (B--E).  First row: stationary Taylor-Green like flow (S-TG) given by Eq. (\ref{eq:stream}).
  Second row: the same as the first row but for the time dependent case (T-TG)
  discussed in Sec. (\ref{sec:timed}).
  }
\label{tab:Rcomparison}
\end{table}
 \subsubsection{Additional exploration during the training}
\label{sssec:exploration}
\begin{figure}[htbp]
\centering
\includegraphics[scale=.8]{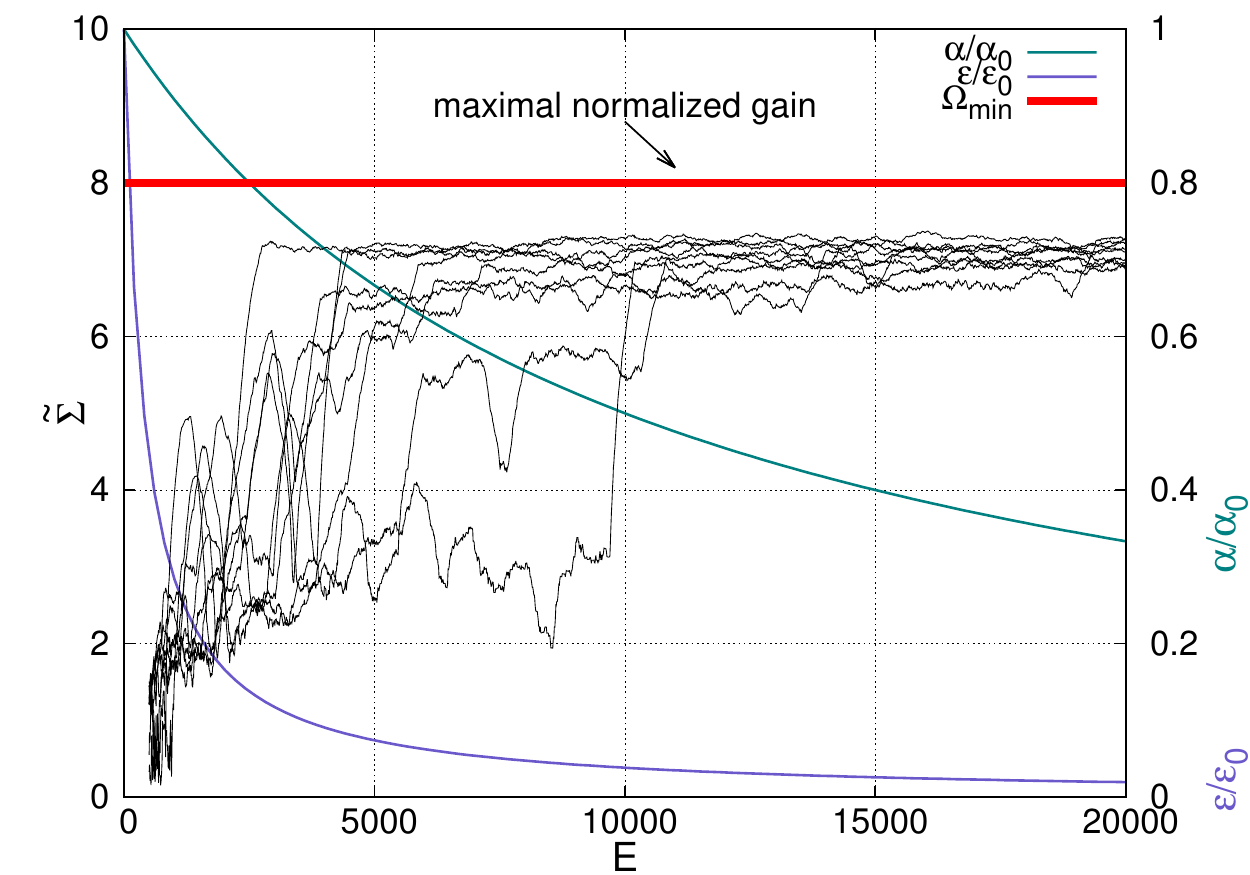}
\caption{
  The normalized learning gain, $\tilde{\Sigma}(E)$ in Eq.~(\ref{eq:normed_tot_return}) against the episodes $E$ for the $\epsilon$-greedy reinforcement learning algorithm. The representation is comparable to Fig.~\ref{fig:Return} (scale on left vertical axis)  .
  The two continuous curves represent
  the adiabatic decreasing of  both the exploration parameter $\epsilon$ and the learning rate $\alpha$ (scale on right vertical axis). }\label{fig:decrease}
\end{figure}

\begin{figure}
\centering
\includegraphics[scale=1,trim={10mm 25mm 10mm 8mm}]{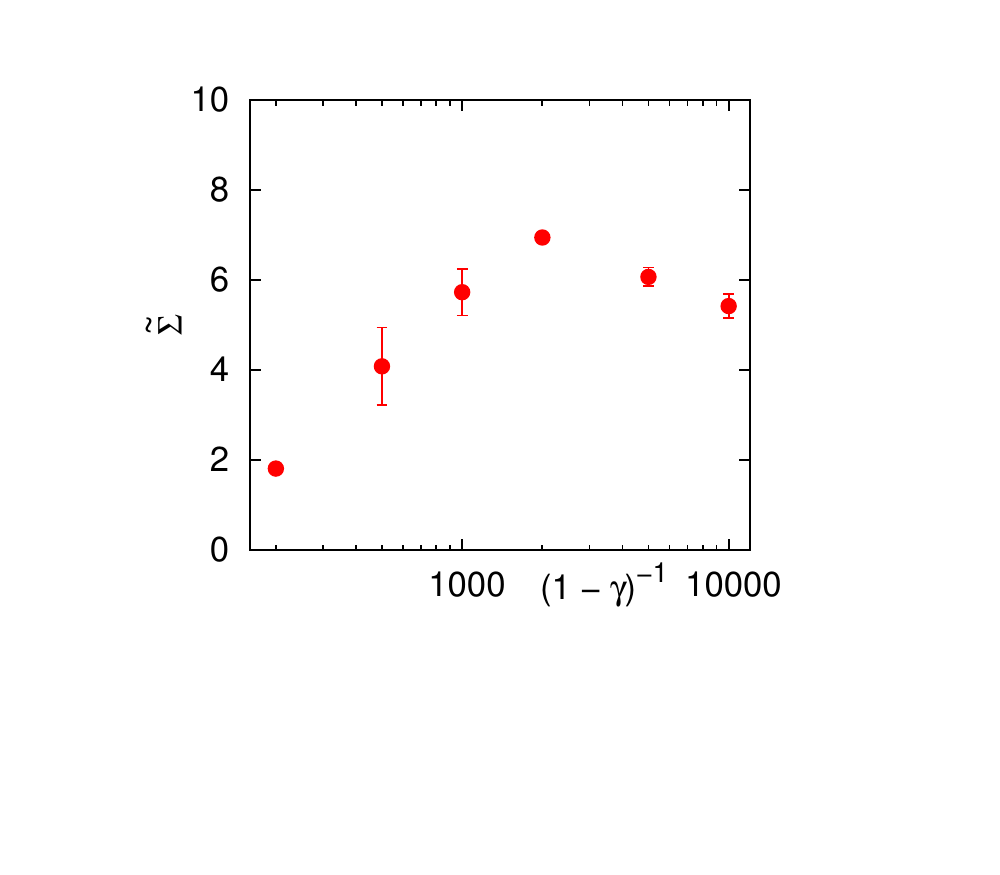}
\caption{The normalized  learning gain $ \tilde{\Sigma}(E) $ as a function of the time horizon of the learning process $ (1-\gamma)^{-1}$ averaged over samples of ten different learning trials.
  Error bars are estimated on the basis of the scatter inside each sample.}\label{fig:rvgamma}
\end{figure}

 In this section we describe how the performance of the reinforcement learning can be improved
 by adding additional exploration during the training session using a non-greedy action ($\epsilon>0$).
 The training phase starts with an initial positive value of $\epsilon$ which is then slowly reduced to zero. This prevents trapping in local
 minima for a transient phase and then slowly the greedy policy is recovered. 
 A particularly simple and efficient scheme is to decrease both
 the learning rate $\alpha$ and the exploration rate $\epsilon$ as a function of the episode, $E$,  during the training:
\begin{equation}
\alpha_E = \alpha_0/(1 + \sigma \, E)\,;\qquad
\epsilon_E = \epsilon_0/(1 + \delta \, E)\,,
\label{eq:adiabatic}
\end{equation}
where $\sigma$ and $\delta$ are positive constants.
In Fig.~\ref{fig:decrease} we show the results for a particular choice,
$\epsilon_0 = 1/1000$; $\alpha_0 = 1/10$; $ \sigma = 1/800$ and $\delta = 1/10000$. These values have been derived by trial and error, rather than by doing a full optimization that goes beyond the scope of the present study. With this selection of parameters the particle have the order of 5000 more exploratory choices during the whole training phase compared to the case of $\epsilon=0$.
We observed that,
by adding additional exploration the smart particles are able to find, in a more systematic way,
approximately optimal policies that are on the same level or better than the best performing policies found in Fig.~\ref{fig:Return}.
The found policies are then stabilized by the adiabatic switch off in Eq.~(\ref{eq:adiabatic}).
It is important to notice that when using the $\epsilon$-greedy method one might need to fine tune the parameters of the learning protocol, $\epsilon_o, \alpha_0,\sigma, \delta$ and that in most cases there is not \textit{a priori} obvious
choice. This is because there exists a trade-off between the advantage of taking locally sub-optimal actions and the risk to drift in the space of possible solutions due to the excess of randomness introduced by the exploration. Similarly, the performance of the algorithm might depend on the sets of allowed states and actions. If there is a too limited number
of options, the particle might not have enough information and/or enough freedom in maneuvering, leading to a failure to reach the target. On the other hand, an excess of inputs and of control might lead to a  slow convergence to the approximately optimal strategy due to the need to explore a
vast number of state-action entries of the $Q$-value matrix.
In Fig.~\ref{fig:rvgamma} we show the sensitivity of our results upon changing the discount factor $\gamma$ but keeping all other training parameters constant. As one can see, the normalized gain remains of the same order when $\gamma$ is varied one order of magnitude.
Only if the discount is too myopic ($\gamma\ll 1$) or too far-sighted ($\gamma\sim 1$) the algorithm fails.
This indicates that the application of the reinforcement learning protocol to the problem of inertial particles studied here is robust to variations in the learning parameters.
Typical combinations of the learning parameters consistently give good rewards and there is no need for fine-tuning of the learning parameters for the application of finding an acceptably good solution. If the application on the other hand is to really find a policy that is arbitrarily close to the approximately optimal behaviour, more care is required in the choice of learning parameters and in the design of the space of allowed states and actions.
\subsubsection{Time-dependent flows}
\label{sec:timed}
\begin{figure}[htbp]
\centering
\includegraphics[scale=.4]{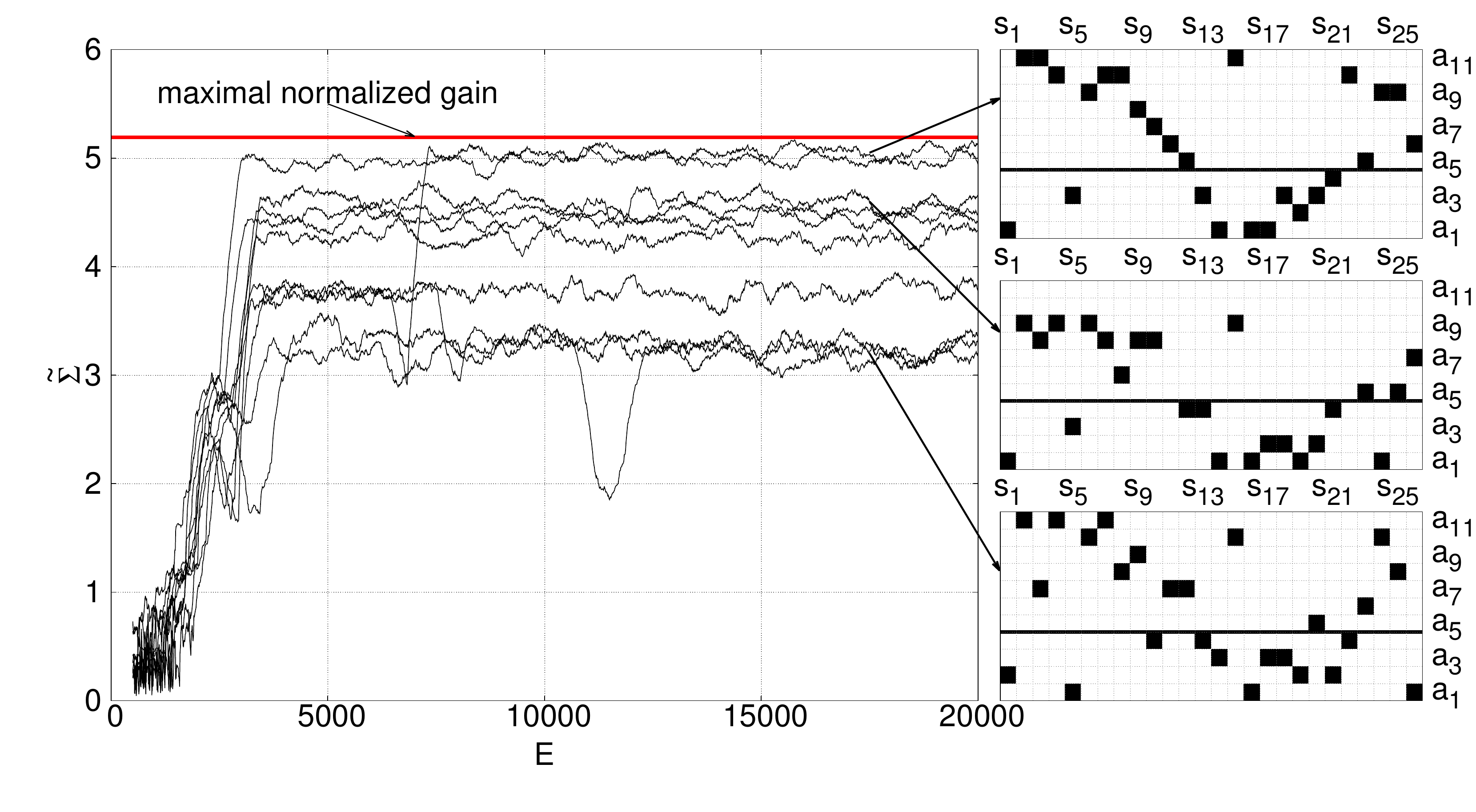}
\caption{ Dependence of the normalized learning gain, $\tilde{\Sigma}$,  for ten different learning processes (black curves) on the two-dimensional time-dependent flow. Every point represents an average over a sliding window of $500$ episodes.
  The three policies shown to the right are for the cases with best, middle and worst final gain. }\label{fig:rtime}
\end{figure}

\begin{figure}[htbp]
\centering
\includegraphics[scale=0.95,trim={20mm 10mm 0mm 0mm}]{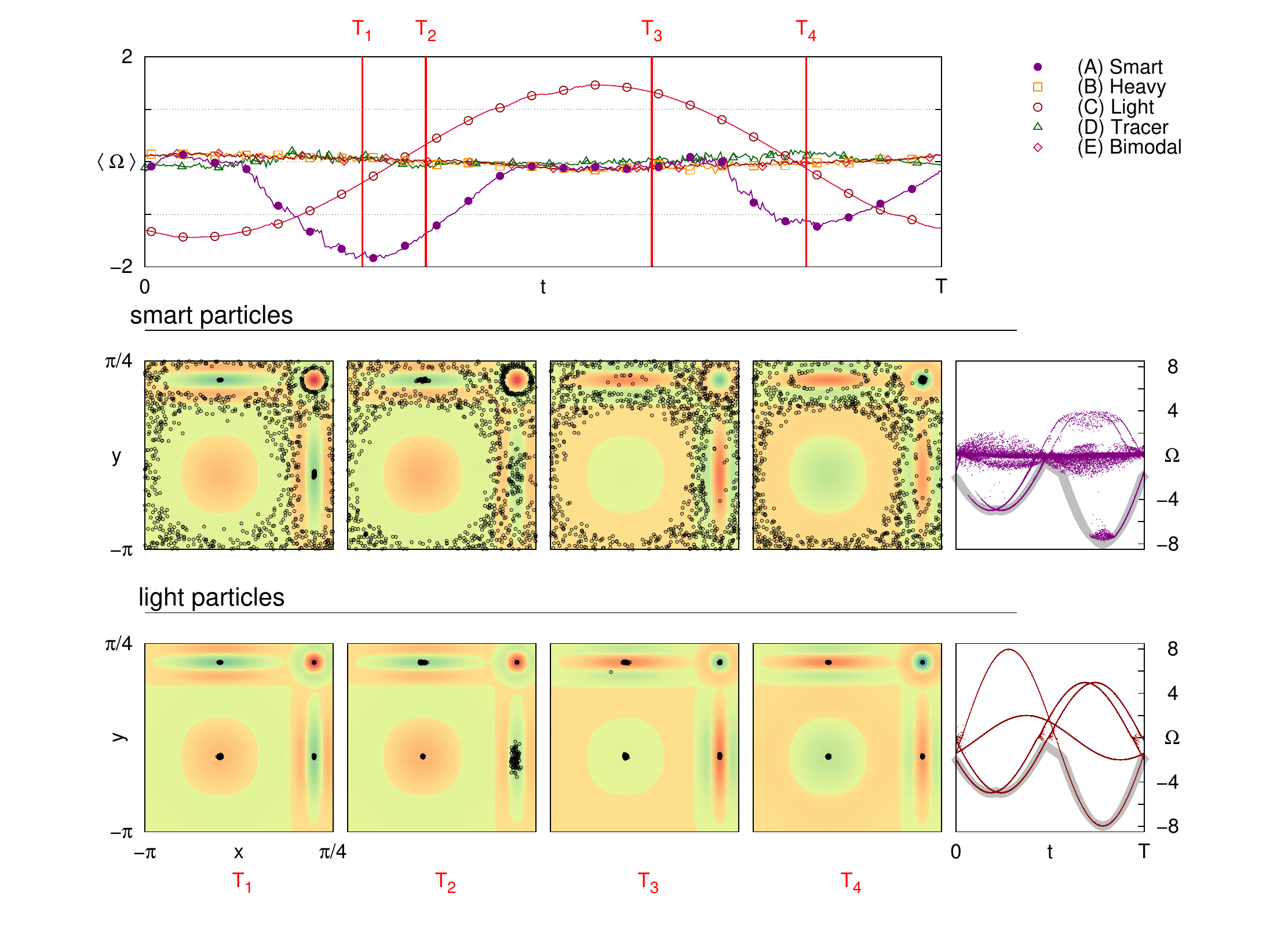}
\caption{
Upper row: average $\Omega$ sampled along the time evolution of $100$ particles for each analyzed type: A-smart (purple), B-heavy (orange), C-light (red), D-tracer (green) and E-particle with bimodal policy (magenta). The average vorticity is periodic with the period $T= 2\pi/\omega_0$ of the underlying flow. \\
Middle row: spatial distribution of $ 100 $ smart particles at the four different times $ T_1, T_2, T_3 $ and $T_4$ highlighted in the top panel.
Lower row: the same but for light particles. The last column on the right shows scatter plots of $ \Omega $ for smart particles (top) and light ones(bottom) throughout the entire period.
The shadowed grey curve represents the instantaneous maximum negative vorticity throughout the flow.
}\label{fig:timedepent}
\end{figure}

It is natural to ask how the approximately optimal policy will perform under
perturbations of the underlying flow or if the algorithm  is
robust when applied to time-dependent and more complex flows.
We have extended the two-dimensional flow in such a way that the
four coefficients $b_1$,\dots,$b_4$ building up the four-vortex flow (see Eq.~(\ref{eq:stream}) in Appendix~I) acquires an out-of-phase oscillating
behavior $ b_i(t) = \cos(\omega_0 t + \phi_i) $, where $ \omega_0 $ is a constant angular frequency
and $ \phi_i $ are four different constant phases. The system has been  trained for the case $ \omega_0 = 0.001$ with constant learning rate
$ \alpha = 0.1 $ and greedy selection of actions ($\epsilon=0$). In Fig.~\ref{fig:rtime} we show the corresponding results
of the total normalized gain during the  training phase,  $\tilde \Sigma (E)$.
Notice that the maximal value of $\tilde \Sigma$ now depends on time, as the instantaneous maximal negative vorticity evolves in time. For our choice of parameters, it turns out that the cube-root of the time average of the
cubed minimal negative vorticity over one oscillation is $(\overline{-\Omega_{\rm min}^3})^{1/3}\approx 5.2$. 
Even in presence of time variations,
the smart particle learns how to move in the flow to follow the most intense negative vortices, which
now oscillates in a non-trivial manner around the four regions of the flow. In Fig.~\ref{fig:rtime} we also show
three approximately optimal policies corresponding to three typical learning events. As one can see,
there are some systematic patterns that are common for all cases. A visual inspection of the spatial distribution of the particles at four different times during one oscillation of the basic flow can be found in Fig.~\ref{fig:timedepent}.
Here it is possible to see how the particles are indeed trying to follow the moving target with high percentage of success. The top panel of Fig.~\ref{fig:timedepent} shows the time dependence of the average vorticity sampled along trajectories of $ 100 $ smart particles that started from random initial positions. The average vorticity is compared with the four reference cases (B-E). 
As one can see, smart particles do not remain oscillating in a confined region but exploit the flow to reach, in average,  more profitable areas.  In the last column on the right of the  same figure we  show the instantaneous distribution of vorticity sampled by the smart particles (top) and by the light case (bottom).
In the figure we also show the instantaneous minimum of vorticity in the two-dimensional spatial configuration.
 We see that while light particles are forced to follow the local extrema of vorticity (positive or negative), the smart particles tend to avoid positive vorticity and to be accumulated in regions of intense negative vorticity, as requested by the reward.

\section{Application to ABC flows}
\label{sec:ABC}
A stationary three-dimensional flow is intriguing because the motion of tracers can be chaotic and very irregular.
The Arnlod-Beltrami-Childress (ABC) flow has been the subject of many studies in turbulence theory.
Its Eulerian velocity field is (in dimensionless coordinates):
\begin{eqnarray}
u_x &=& C \cos y + A \sin z,\nonumber\\
u_y &=& A \cos z + B \sin x,\nonumber\\
u_z &=& B \cos x + C \sin y\,.
\end{eqnarray}
The flow is characterized by three parameters $A$, $B$, and $C$.

Numerical simulations and theoretical arguments show that the ABC flow has tube-like regions in space within which the streamlines of the flow are confined and the velocity is essentially one-dimensional~\cite{dombre1986chaotic}.
Since vorticity $\ve\Omega\equiv\ve\nabla\wedge\ve u$ is parallel to the velocity in an ABC flow, $\ve\Omega=\ve u/2$, these tubes are referred to as principal vortices.
Because of symmetries, an ABC flow with $A=B=C$ has three pairs of principal vortices that are mainly aligned with the three directions $\hat x$, $\hat y$ and $\hat z$.
Each pair consists of two vortices of opposite sign of velocity and vorticity.
Within the principal vortices of an ABC flow the dynamics is regular, while outside it may become chaotic.
Similar to the two-dimensional flow in Section~\ref{sec:TGflow}, we impose here as a target for the smart particle to maximize the magnitude of its vorticity $\Omega\equiv|\ve\Omega|$.
In order to achieve this goal, the particle needs to navigate a complex flow landscape to target the principal vortices with maximal vorticity.

\subsection{Algorithm implementation}
\label{sub:Ai}
To show how general the success of the reinforcement learning is, we adopt a slightly modified version of the learning framework implemented in Section~\ref{sec:TGflow}.
We keep ${\rm St}=0.2$ fixed, and use as an action to change the value of $\beta$. Allowed values are equally distributed in $N_a$ levels between $0$ and $3$.
The state of the particle is given by either $|\ve\Omega|$ or one of the components of $\ve\Omega$, equally partitioned in $N_s$ levels between the minimal and maximal value that can be obtained in the ABC flow.
We use as reward $|\ve\Omega|^3$ averaged over the time the particle spend between state changes. This is in contrast to Section~\ref{sec:TGflow}, where the cubed vorticity was evaluated at the position of the state change.
As in Sec.~\ref{sec:TGflow} we use optimistic learning, where the entries of the initial $Q$-value matrix is $N$ times the maximal reward and $N=1000$ is the number of state changes per episode.
We keep the learning rate fixed, $\alpha=0.1$, use a greedy policy, $\epsilon=0$, use a discount factor $\gamma=0.97$ and use no noise in Eq.~(\ref{eq:model}), $\chi=0$.\\

\subsection{Flow parameters}
For the ABC flow we use light particles ($\beta=3$) as naive reference particles.
Principal vortices are traps for light particles: depending on the initial condition a light particle ends up in either one of the principal vortices.
In a symmetric ABC flow ($A=B=C$) the average magnitude of vorticity along trajectories in each of the principal vortices is identical and the smart particle only marginally manages to outperform a light particle (not shown).
We therefore consider ABC flows with weakly broken symmetry, $2A=B=C=1$, and with strongly broken symmetry, $4A=2B=C=1$.
In the weakly broken case, the dynamics in one direction is distinct from the other two, and in the strongly broken case all three directions have different dynamics.
For the case of an asymmetric ABC flow, the situation is similar to the four-vortex flow studied above.
If the particle is constantly light, it will be attracted to a vortex region depending on its initial condition.
However, since different principal vortices have different intensity of vorticity in the asymmetric ABC flow, not all light particles will end up in the region of strongest vorticity (similar to the four-vortex flow above, where light particles end up in either of the vortices depending on its initial condition).
Giving the smart particle appropriate information about the flow, we expect it to be able to learn to go to the principal vortices of highest vorticity and consequently beat the light particle.


\subsection{Results}
\begin{figure*}[t]
\begin{overpic}[width=5.5cm,clip]{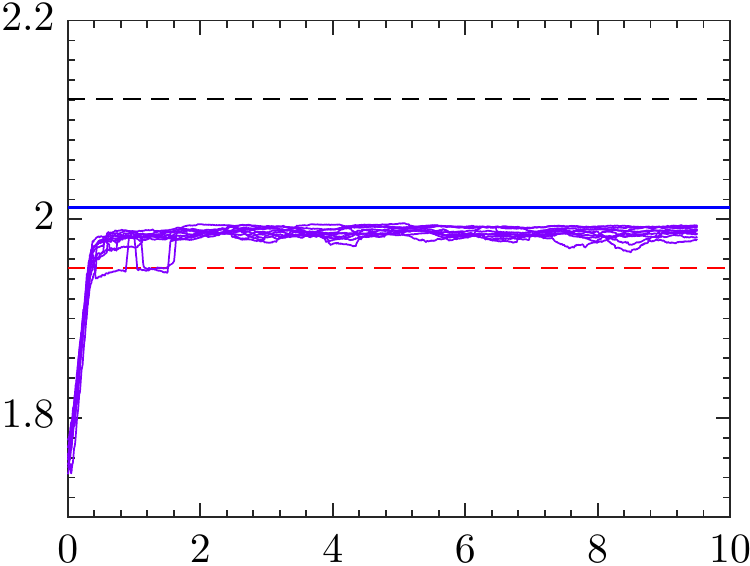}
\put(87,11){{{\bf a}}}
\put(45,-6){{{$E$ }}}
\put(82,-6){{{$\times$ 1000}}}
\put(-7,40){\rotatebox{90}{$\tilde \Sigma$}}
\end{overpic}
\hspace{2mm}
\begin{overpic}[width=5.45cm,clip]{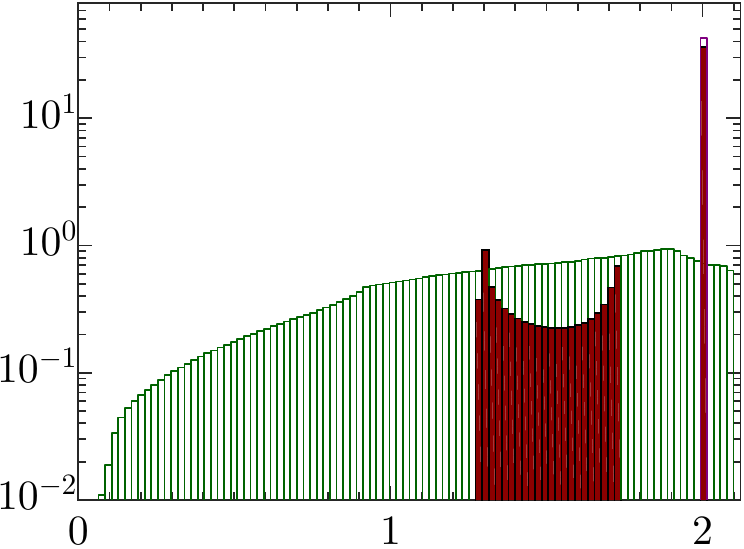}
\put(87,11){{{\bf b}}}
\put(55,-6){$|\ve\Omega|$}
\put(-7,45){\rotatebox{90}{$P(|\ve\Omega|)$}}
\end{overpic}
\hspace{2mm}
\begin{overpic}[width=5.15cm,clip]{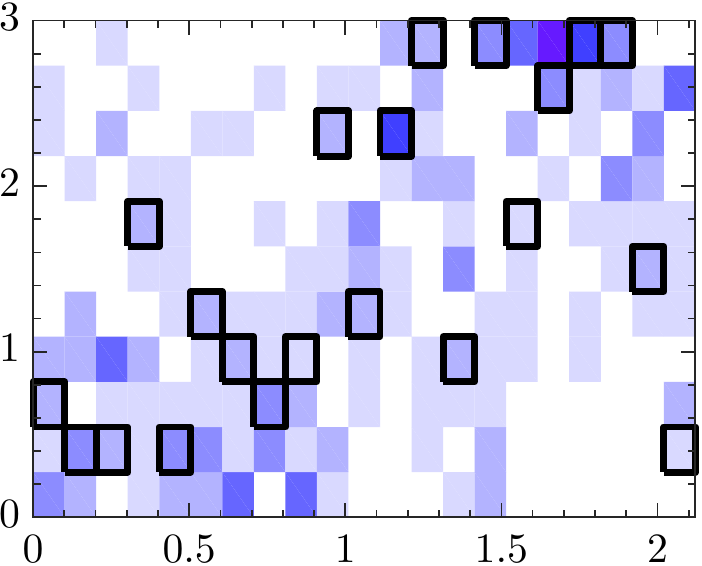}
\put(87,11){{{\bf c}}}
\put(55,-6){$|\ve\Omega|$}
\put(-7,45){\rotatebox{90}{$\beta$}}
\end{overpic}
\raisebox{3mm}{
\begin{overpic}[width=0.5cm,height=3.75cm,clip]{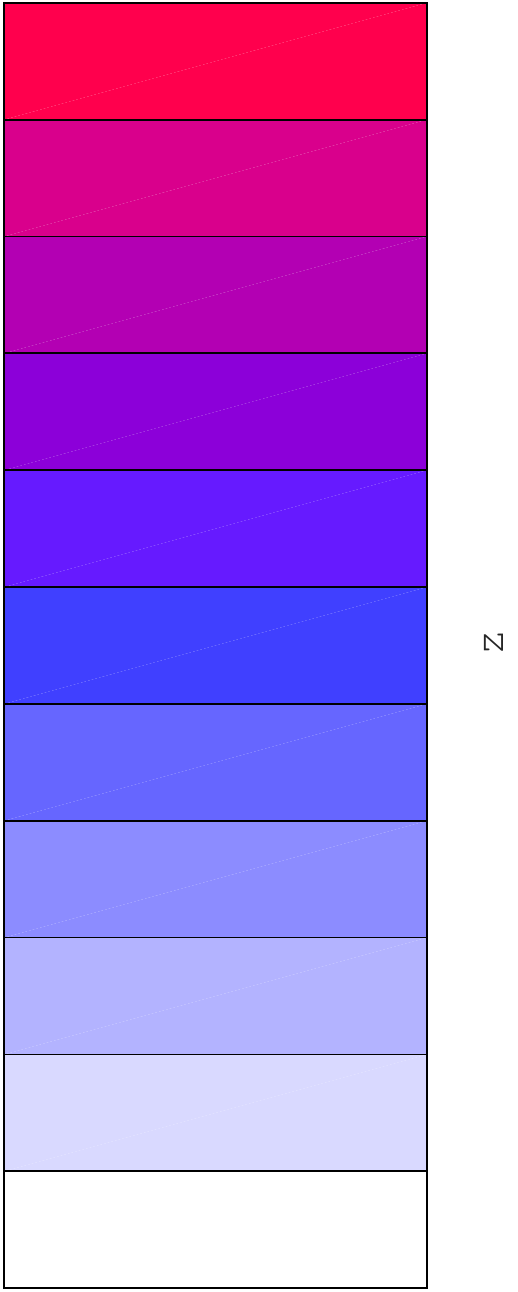}
\put(13,-2){{0}}
\put(13,97){{10}}
\put(15,68){\rotatebox{-90}{Frequency}}
\end{overpic}
}
\caption{Result from training of particles in ABC flow with weakly broken symmetry ($2A=B=C=1$).
{\bf a} Normalized total gain $\tilde \Sigma =\langle|\ve\Omega|^3\rangle^{1/3}$ [Eq.~(\ref{eq:normed_tot_return})] as a function of episode $E$.
The maximal vorticity in the flow is shown as black-dashed.
Red-dashed line shows the steady-state average 
$ <|{\bf \Omega}|^3>^{1/3}_\infty $ along the trajectories of many naive light particles.
Solid blue line shows the corresponding average $<|{\bf \Omega}|^3>^{1/3}_\infty$ using the best policy obtained. Black dashed line shows maximal vorticity.
{\bf b} Distribution of magnitude of vorticity $|\ve\Omega|$ for smart particles (blue peak), naive particles with $\beta=1$ (green) and naive particles with $\beta=3$ (red region and peak).
{\bf c} Frequency of optimal action for each state using the ten final policies for the cases displayed in panel {\bf a}.
Black boxes highlight the best policy in panel {\bf a}.
}\label{fig:training_ABC}
\end{figure*}
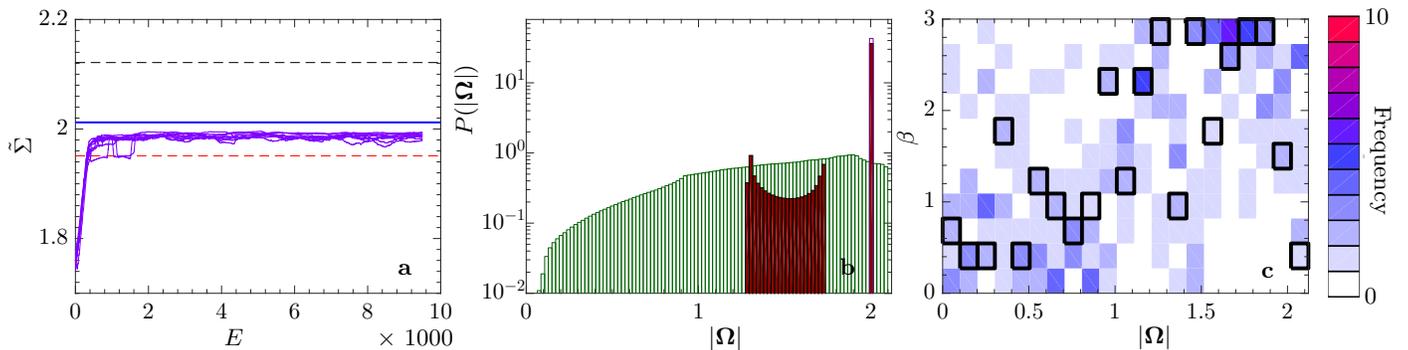

Figure~\ref{fig:training_ABC}{\bf a} shows the evolution of the normalized total gain $\tilde \Sigma$ for training of particles in an ABC flow with weakly broken symmetry.
The training has been performed using $|\ve\Omega|$ as the state and repeated ten times.
The resulting curves (solid purple) are compared to the maximal vorticity in the flow (black dashed) and the steady-state average $\langle|\ve\Omega|^3\rangle_\infty^{1/3}$ for naive particles with constant $\beta=3$ (red dashed).
We remark that the value of the purple curves is lower than the steady state average due to the initial transient before the steady state is reached.
We therefore also plot as solid blue the steady state average $\langle{|\ve\Omega|^3}\rangle_\infty^{1/3}$ for the best policy (highest averaged reward at the last episode) for the data in Fig.~\ref{fig:training_ABC}{\bf a}.
Fig.~\ref{fig:training_ABC}{\bf b} shows the steady-state distribution of the magnitude of vorticity for the best policy in Fig.~\ref{fig:training_ABC}{\bf a} (blue peak),  for tracer particles (green solid line), and for naive light particles (red line and peak).
The distribution for the tracer particles shows that over the entire flow, vorticity is more or less uniformly distributed with some reduced probability at small values.
The distribution for the smart particles instead shows a sharp peak.
A similar peak is also found for the light particles, but these also show a band of vorticities around $|\Omega|=1.5$.
Fig.~\ref{fig:training_ABC}{\bf c} shows, for the cases in Fig.~\ref{fig:training_ABC}{\bf a}, the best policy found and the frequency at which the ten final policies select actions for each state.

For non-small values of ${\rm St}$ (${\rm St}=0.5$ and ${\rm St}=2$) the reinforced learning scheme basically finds the naive solution, i.e. to be heavy if $|\ve\Omega|$ is small, and light if $|\ve\Omega|$ larger than some threshold value (not shown).
These solutions perform at the same level as the naive solution of being light with constant $\beta=3$.
For the case ${\rm St}=0.2$ considered here, the smart particle find a qualitatively different solution that beats the naive solution, although the final gain is only approximately 5\% better than the naive solution.
As shown in Fig.~\ref{fig:training_ABC}{\bf b} the distribution of vorticity for the smart particle and the naive light particle are similar but with one important difference.
The sharp peak close to $|\ve\Omega|=2$ corresponds to the principal vortices in the $z$-direction. 
While all initial conditions end up in these vortices for the smart particles, some initial conditions for naive particles ends up in the subdominant principal vortices orthogonal to the $z$ direction, leading to the band of vorticities around $|\Omega|=1.5$ which explains why the naive particles have a lower gain.
As shown in Fig.~\ref{fig:training_ABC}{\bf c}, the trend in the policy is basically to be light when vorticity is high, and heavy when vorticity is low, but there also seems to be some structure needed in the intermediate vorticity states that enables the smart particles to outperform the naive one.

\begin{figure*}[htbp]
\begin{overpic}[width=0.41\textwidth,clip]{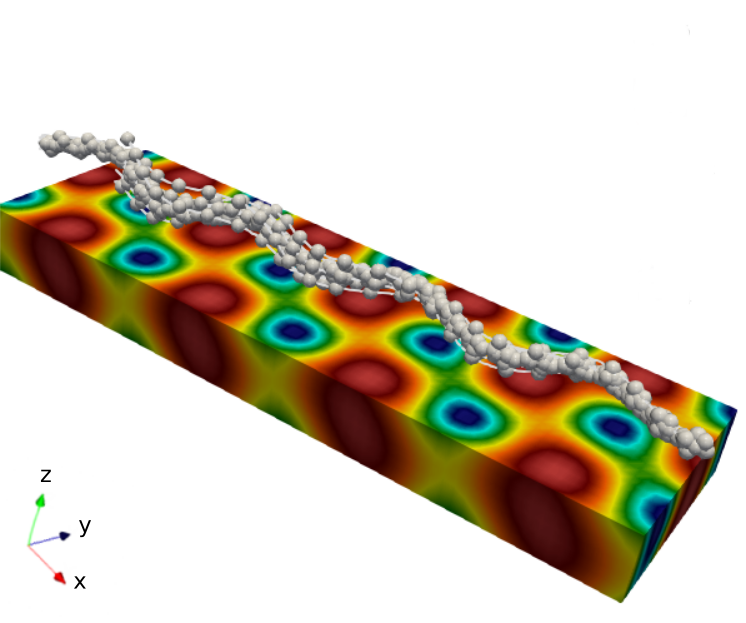}
\put(5,73){{{\bf a}}}
\end{overpic}
\begin{overpic}[width=0.49\textwidth,clip]{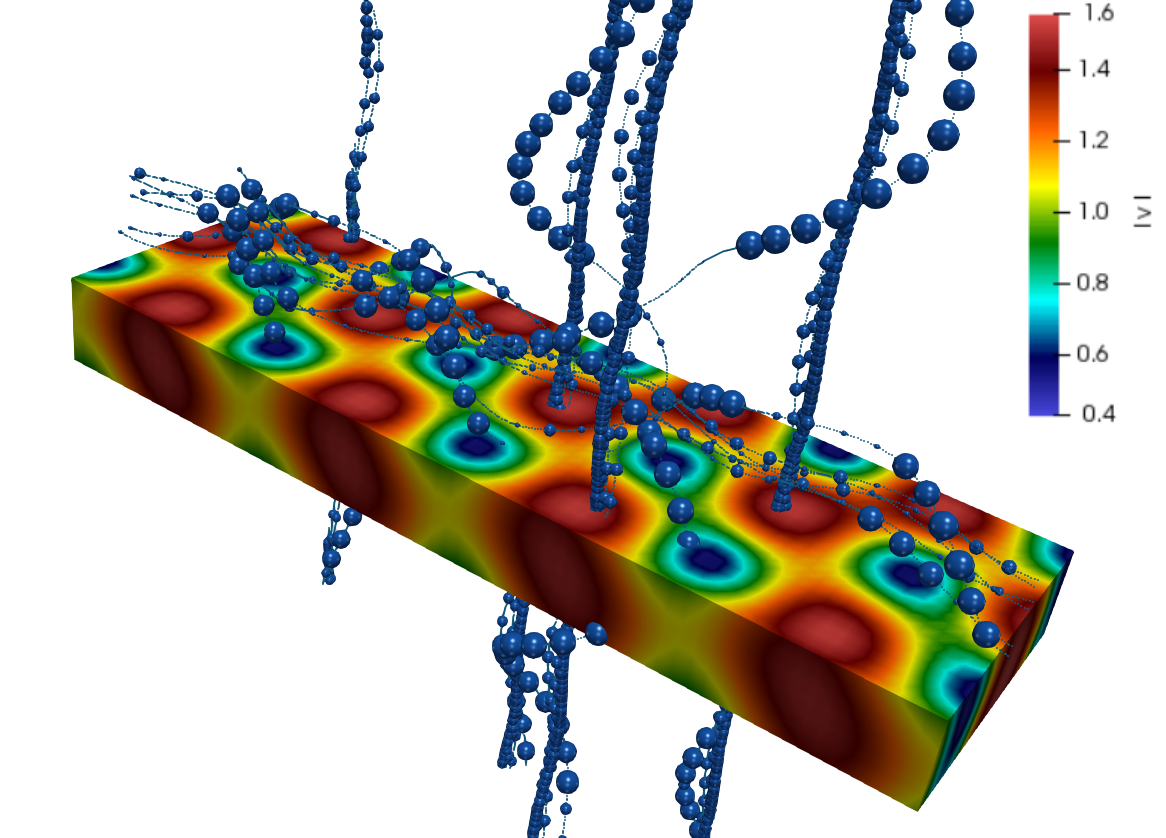}
\put(5,61){{{\bf b}}}
\end{overpic}	
\caption{
Particle trajectories in an ABC flow ($4A=2B=C=1$) starting from an identical initial condition of a naive light particle ({\bf a}) and a smart particle {\bf b}.
The trajectories show that a smart particle  manages to find the optimal vertical principal vortices independent of the initial condition, while the naive light particle sometimes get stuck in suboptimal horizontal principal vortices. The plotted particle sizes are proportional to the value of $\beta$.
The vorticity field is color-coded. To allow the visualization of particle trajectories we have represented only a slice of it.
}\label{fig:ABC_Trajs}
\end{figure*}

For the ABC flow with strongly broken symmetry, light particles distribute on two pairs of principal vortices, with roughly $80\%$ of the particles in the strong principal vortices in the $\hat z$-direction and the rest on the weaker principal vortices in the $\hat x$-direction. One such trajectory is shown in Fig.~\ref{fig:ABC_Trajs}{\bf a}.
Training of the smart particle on the other hand, using $\Omega_z$ as the state, allows it to find strategies that target the dominant principal vortices in the $\hat z$ direction for all tested initial conditions.
One such example is shown in Fig.~\ref{fig:ABC_Trajs}{\bf b}: starting from the same initial condition the smart particle reaches the optimal vortex, while the naive particle ends up in a subdominant vortex.

\begin{figure*}[htbp]
\begin{overpic}[width=4cm,clip]{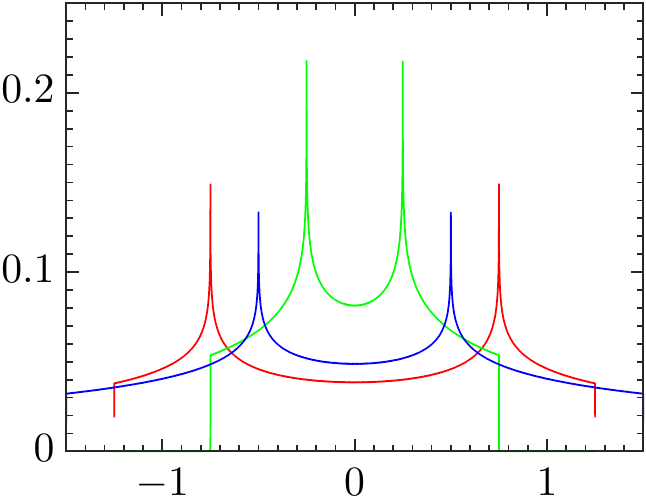}
\put(52,-8){$\Omega_i$}
\put(-8,35){\rotatebox{90}{$P(\Omega_i)$}}
\put(59,68){$\Omega_y$}
\put(67,45){$\Omega_z$}
\put(75,55){$\Omega_x$}
\end{overpic}
\caption{Flow distribution of different components of $\Omega$: $\Omega_x$ (red), $\Omega_y$ (green), $\Omega_z$ (blue) for an ABC flow with strongly broken symmetry, $4A=2B=C=1$.
}\label{fig:flowdistribution}
\end{figure*}

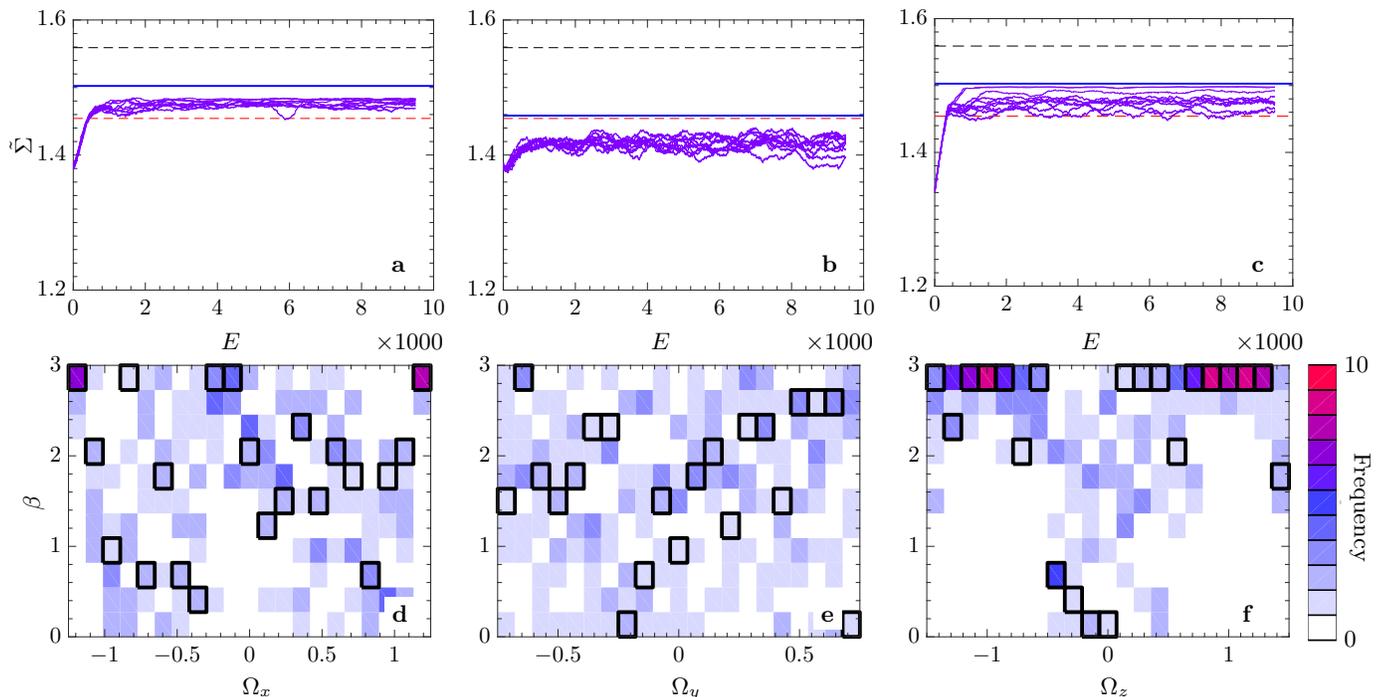
\begin{figure*}[t]
\begin{overpic}[width=5.4cm,clip]{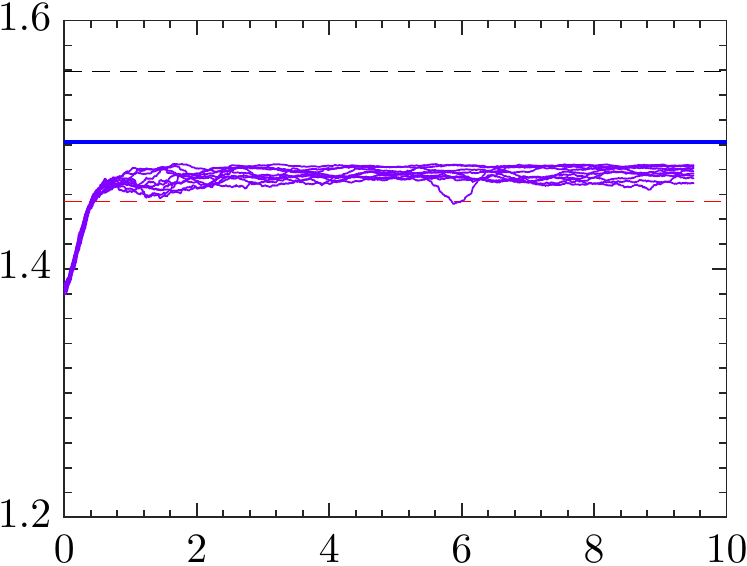}
\put(87,11){{{\bf a}}}
\put(-7,40){\rotatebox{90}{$\tilde \Sigma$}}
\put(45,-8){{{$E$ }}}
\put(83,-8){{{$\times$1000}}}
\end{overpic}
\hspace{0.1cm}
\begin{overpic}[width=5.4cm,clip]{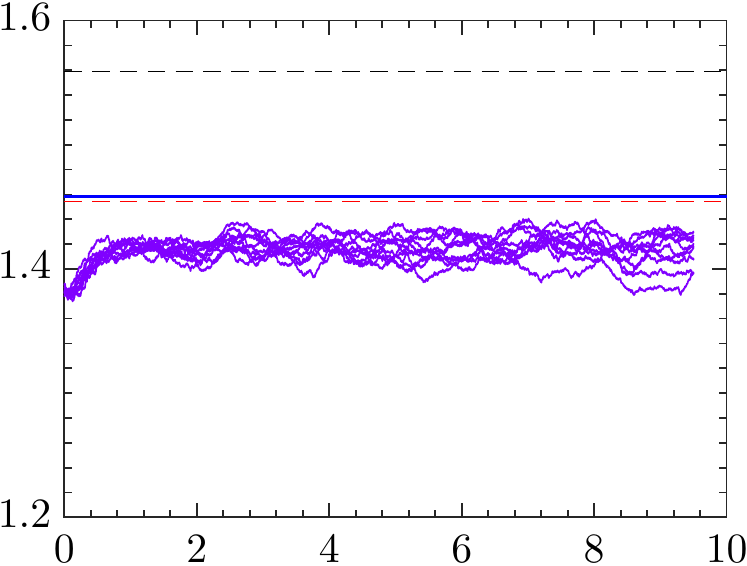}
\put(87,11){{{\bf b}}}
\put(45,-8){{{$E$ }}}
\put(83,-8){{{$\times$1000}}}
\end{overpic}
\hspace{0.1cm}
\begin{overpic}[width=5.4cm,clip]{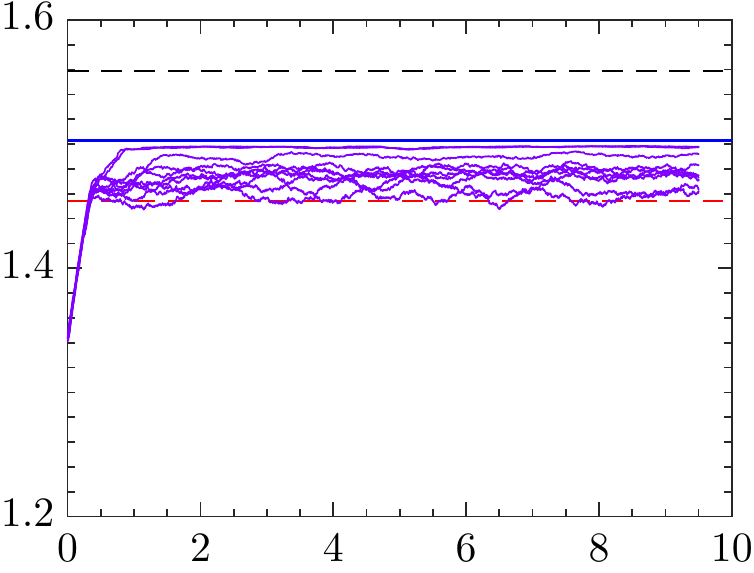}
\put(87,11){{{\bf c}}}
\put(45,-8){{{$E$ }}}
\put(83,-8){{{$\times$1000}}}
\end{overpic}
\hspace{2cm}
\\
\vspace{0.5cm}
\hspace{0.55cm}
\begin{overpic}[width=5.1cm,clip]{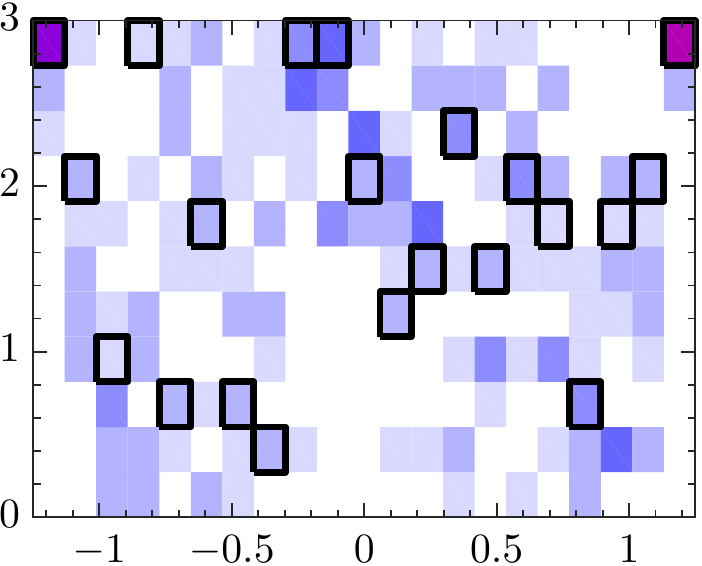}
\put(87,11){{\colorbox{white}{{\bf d}}}}
\put(50,-8){$\Omega_x$}
\put(-7,40){\rotatebox{90}{$\beta$}}
\end{overpic}
\hspace{0.4cm}
\begin{overpic}[width=5.1cm,clip]{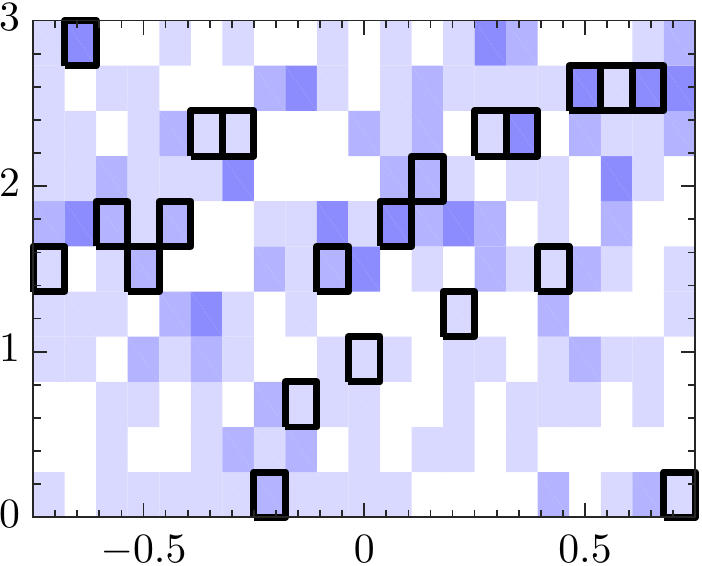}
\put(87,11){{\colorbox{white}{{\bf e}}}}
\put(50,-8){$\Omega_y$}
\end{overpic}
\hspace{0.4cm}
\begin{overpic}[width=5.1cm,clip]{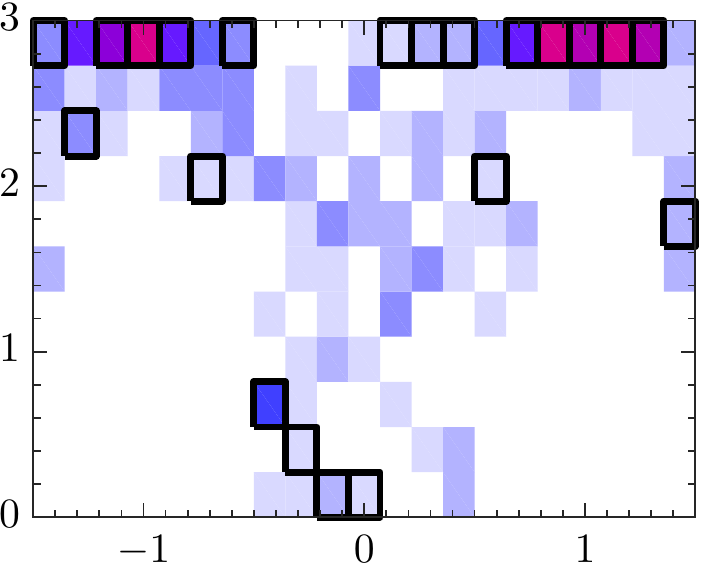}
\put(87,11){{{\bf f}}}
\put(50,-8){$\Omega_z$}
\end{overpic}
\raisebox{0.3cm}{
\begin{overpic}[width=0.45cm,height=3.68cm,clip]{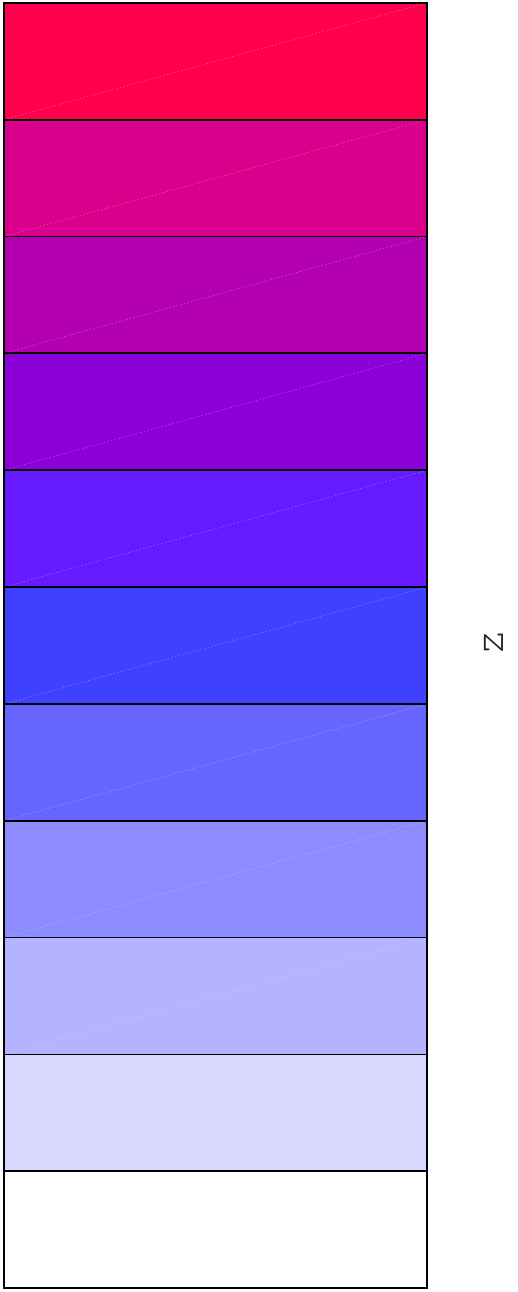}
\put(13,-2){{0}}
\put(13,97){{10}}
\put(15,68){\rotatebox{-90}{Frequency}}
\end{overpic}
}
\caption{Result from training of particles in an ABC flow with strongly broken symmetry ($4A=2B=C=1$).
Data is displayed in the same manner as Fig.~\ref{fig:training_ABC}.
{\bf a}--{\bf c} Normalized total gain $\tilde \Sigma (E)$ as a function of episode during training using the components $\Omega_x$ ({\bf a}), $\Omega_y$ ({\bf b}), $\Omega_z$ ({\bf c}) as state.
{\bf d}--{\bf f}
Corresponding frequency of optimal action for the final policies in panels {\bf a}--{\bf c}.
}\label{fig:training_ABC_asymmetric}
\end{figure*}

In Fig.~\ref{fig:ABC_Trajs} we observed that the smart particles using $\Omega_z$ as state recognizes the two principal vortices of highest vorticity, and therefore get a higher reward than the light particles with constant $\beta = 3$.
The training progress and resulting policy is shown in Fig.~\ref{fig:training_ABC_asymmetric}{\bf c},{\bf f}.
The structure of the $Q$-value matrix suggests that the policy used is quite simple: be
heavy ($\beta=0$) when $|\Omega_z|$ is smaller than some threshold and light ($\beta=3$) otherwise. This observation is supported
by the data in Table~\ref{tab:naiveguess}. The normalized returns quoted in Table~\ref{tab:naiveguess} show that if the
threshold is chosen around $N^{*} = 7 \sim 9$, where $N^{*}$  is the number of states where the particle is heavy, the bimodal policy in Table~\ref{tab:naiveguess} performs roughly at the same level as the trained solution. This value of the threshold (also found by the smart particle) does not follow immediately from the distribution of vorticity, shown in Fig.~\ref{fig:flowdistribution}.
\begin{table}
\begin{tabular}{l | l l l l l l l l l l l}
$N^{*}$ &19&17&15&13&11&9&7&5&3&1&0\\
$\tilde R_{\rm tot}$ &1.01&1.04&1.13&1.19&1.49&1.50&1.50&1.49&1.45&1.44&1.45
\end{tabular}\caption{Normalized return $\tilde R_{\rm tot}$ defined in (\ref{eq:norm_disc_ret}) for the examination phase with $Q$-value matrix such that the optimal action $a^{*}(s)$ is $0$ (heavy) for a number $N^{*}$ of centered states and $3$ (light) for the remaining states. As an example, $N^{*} = 9$ and $N_{s}=21$ states used here gives $a^{*}(s)= \{3,3,3,3,3,3,0,0,0,0,0,0,0,0,0,3,3,3,3,3,3\}$.
\label{tab:naiveguess}
}
\end{table}

Fig.~\ref{fig:training_ABC_asymmetric}{\bf a},{\bf d} and Fig.~\ref{fig:training_ABC_asymmetric}{\bf b},{\bf e} show results from training with $\Omega_x$ and $\Omega_y$ as states.
We find that if $\Omega_x$ is used as the state, the smart particle learns to outperform the naive light particle, using a non-trivial strategy (Fig.~\ref{fig:training_ABC_asymmetric}{\bf d}).
If $\Omega_y$ is used as the state on the other hand, the smart particle is not able to find a strategy that outperforms a light particle.
This shows that it is important that the smart particle must measure appropriate information about the flow to be able to find a good strategy.
Fig.~\ref{fig:flowdistribution} shows the distribution of the components in the underlying ABC flow.
The distribution of $\Omega_y$ is narrower than the distributions of $\Omega_x$ and $\Omega_z$. This explains why it is hard to use $\Omega_y$ as the state in order to find regions with large $|\ve\Omega|$.

\subsection{Evaluation of the algorithm}
\begin{figure}[htbp]
\begin{overpic}[width=0.8\textwidth,clip]{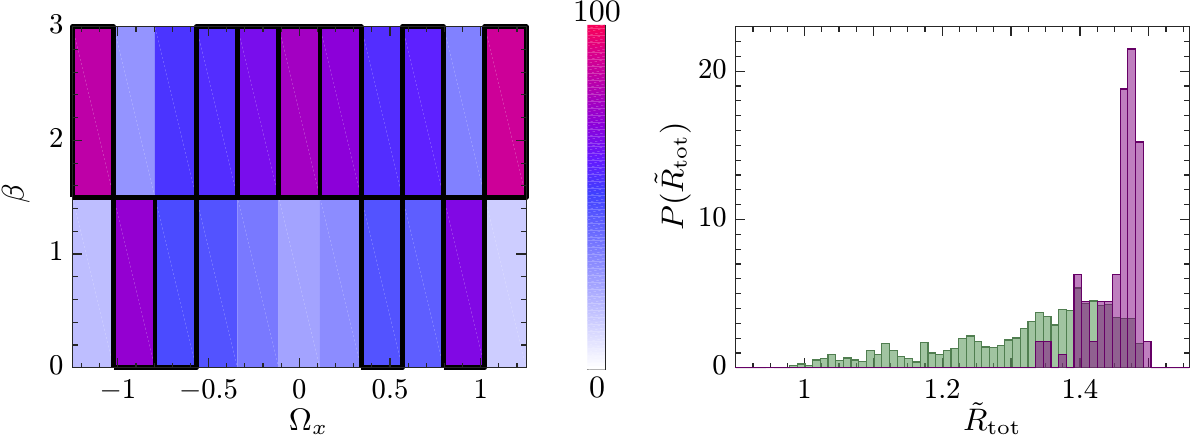}
\put(6.5,31){\colorbox{white}{${\bf a}$}}
\put(64,31){${\bf b}$}
\end{overpic}
\caption{
Evaluation of the performance of smart particles in an ABC flow ($4A=2B=C=1$) with $N_a=2$ actions ($\beta=0$ or $\beta=3$) and $\Omega_x$ divided into $N_s=11$ states.
{\bf a} Solid black lines indicate the best policy obtained by the smart particles.
This policy coincides with the global optimal policy for the set of states and actions used.
Colors indicate frequency of the optimal action for each state based on $100$ training sessions.
{\bf b} Green bars show distribution of average normalized return $\tilde R_{\rm tot}$ for all $N_s^{N_a}=2048$ possible $Q$-value matrices.
Purple bars show distribution of $\tilde R_{\rm tot}$ for the 100 training sessions.
}\label{fig:ABC_evaluation}
\end{figure}

In general it is hard to evaluate the success of the reinforcement learning algorithm because the global optimal policy is not known.
Due to the vast size of possible $Q$-value matrices it is in general not possible to do a brute force approach, by testing all possible $Q$-value matrices.
However, in the current problem it turns out that the algorithm is able to find non-trivial solutions also if the number of actions and number of states are reduced.
We consider training using $\Omega_x$ as a state (the case in Fig.~\ref{fig:training_ABC_asymmetric}{\bf a},{\bf d}), with a reduced number of actions $N_a=2$ ($\beta$ can take the values $0$ or $3$) and states $N_s=11$.
The results for $100$ training sessions are displayed in Fig.~\ref{fig:ABC_evaluation}{\bf a}.
The general trend is non-trivial, the particle should be light for large $|\Omega_x|$, mainly heavy in a range of intermediate $|\Omega_x|$ and light again for small $\Omega_x$.

There are $2^{11}=2048$ possible $Q$-value matrices.
We evaluate the normalized total return $\tilde R_{\rm tot}$ for each $Q$-value matrix averaging over $1000$ episodes starting from $1000$ predetermined initial conditions that are identical for each tested $Q$-value matrix.
Fig.~\ref{fig:ABC_evaluation}{\bf b} shows the distribution of $\tilde R_{\rm tot}$ for all the $2048$ possible $Q$-value matrices (green).
Also displayed is the distribution of $\tilde R_{\rm tot}$ obtained from the $100$ policies underlying Fig.~\ref{fig:ABC_evaluation}{\bf a} (purple).
We find that the policies obtained by reinforcement learning in general lies close to the global optimal solution, and that in some instances the true global optimum is reached.
The optimal normalized return is $\tilde R_{\rm tot}=1.49$, which is comparable to the best gain $\tilde \Sigma=1.50$ for the case $N_a=11$ and $N_s=21$ (the data in Fig.~\ref{fig:training_ABC_asymmetric}{\bf a} and {\bf d}), but the case with fewer states and actions is more likely to get stuck at poor solutions with a lower return.


\section{Conclusions}
\label{sec:co}
In this paper, we have shown how smart inertial particles can learn to sample the most intense vortical structures in fluid flows of different complexity: a two-dimensional stationary Taylor-Green like configuration, a two-dimensional time-dependent Taylor-Green like flow and three-dimensional ABC flows.
We achieve this goal by defining the problem within the reinforcement learning framework and using the so-called one-step $Q$-learning algorithm to obtain approximally optimal policies in an iterative way. We evaluate the learning performance by comparing the acquired ability of smart particles to reach the target region with respect to the analogous skill of particles with fixed size. For the 2D examined flows, it is found that the learnt policies allow outperforming of particles that cannot modulate their size or that can only do it rudimentary. While the trajectories of smart particles have high density in the target region, this region is never sampled, or just rarely, in all the other instances. 
Even better strategies are obtained by adopting an $\epsilon$-greedy algorithm, i.e. allowing additional exploration during the training session and letting the learning parameter $ \alpha $ decrease in time to stabilize the found policies. Smart particles tend to elude positive vorticity and to accumulate in regions of intense negative vorticity even in the time-dependent flow. For the  investigated three-dimensional flows smart particles mainly outperform fixed-size particles except for the case of particles with non-small Stokes numbers in the slightly asymmetric ABC flow, for which the smart particles perform on the same level as naive light particles. 
In this context it emerges more clearly how general the success of the reinforcement learning is, but also that achieving the predetermined goal depends on the properties particles can measure and on how much the target can be discriminated from non-interesting regions.
Despite that a fully realistic description of the particle dynamics and the actual complexity of real flows is far from the goal of this paper, we provide a proof of concept for the possibility to engineer smart inertial particles and to make a case for the use of reinforcement learning algorithms for this purpose. Similar control strategies might also be implemented with heavy particles subjected only to the Stokes drag as would be the case for diatom algae~\cite{sarthou2005diatoms}. 
It is important to stress that the point particle approximation is only valid if the particle is small enough and that the approximation of one way coupling is only valid if the particle does not change state to frequently in physical time.
There is room for improvement in many directions.
For  instance, other ways to control the dynamics of engineered particles could be  by changing their chirality or ellipsoidal structures. Moreover, other sensory inputs could be explored,  one example being the temperature in convection.\\
{\sc Acknowledgments}  L.B., K.G. and S.C. acknowledge funding from the European
Research Council under the European Union’s Seventh
Framework Programme, ERC Grant Agreement No. 339032. K.G. acknowledges funding from the Knut and Alice Wallenberg Foundation, Dnar. KAW 2014.0048.
\section{Appendix I}
The flow domain is a square of size $ L = 5\pi/4$ consisting of four quadrilaterals, (two squares and two rectangles) of sides $L_1 = \pi $ and $L_2 = \pi/4 $, ($L = L_1 + L_2$). 
The velocity and vorticity fields are built from the following  stream function, made out of the superposition of four
different vorticity blobs placed at the center of each subdomain: 
\begin{eqnarray}\label{eq:stream}
\psi(x,y) &=& b_1 G_{1}(x) G_{1}(y) \sin \biggl(\frac{x\pi}{L_1}\biggr)\sin\biggl(\frac{y\pi}{L_1}\biggr) P(x)P(y)\nonumber \\
&+& b_2 G_{2}(x) G_{1}(y) \sin\biggl(\frac{x\pi}{L_2}\biggr)\sin\biggl(\frac{y\pi}{L_1}\biggr)P(y)\nonumber \\
&+& b_3 G_{2}(x) G_{2}(y) \sin\biggl(\frac{x\pi}{L_2}\biggr)\sin\biggl(\frac{y\pi}{L_2}\biggr) \nonumber \\
&+& b_4 G_{1}(x) G_{2}(y) \sin \biggl(\frac{x\pi}{L_1}\biggr)\sin\biggl(\frac{y\pi}{L_2}\biggr) P(x),
\end{eqnarray}
where
\begin{align}
G_{1}(x) &= \exp(-(x-\bar{x}_{1})^2/(2\Delta_{1}^{2})), \nonumber\\
G_{2}(x) &= \exp(-(x-\bar{x}_{2})^2/(2\Delta_{2}^{2})), \nonumber
\end{align}
are the Gaussian functions that modulate the vortical structures with widths $\Delta_1 = L_1/4$, $\Delta_2=L_2/4$ and centers in $\overline{x}_1 = -L_1/2$, $ \overline{x}_2 = L_2/2 $
and $$
P(x) = [x - \bar{x}_{1} - (L - L_1/2)][x - \bar{x}_{1} + (L - L_1/2)]
$$

is a polynomial of degree $2$ such that the orthogonal velocity component vanishes at the boundaries of the domain. 
The  coefficients, $b_i$ are fixed as $(b_1,b_2,b_3,b_4)=(-0.1,0.02,-0.10,0-02)$ and determine the intensity of the vortical structures to be (approximately) $(5, -8, 5, -2)$ (see Fig.~\ref{fig:vorticity}).
Reflecting boundary conditions are used to confine the particles inside the volume.
\bibliography{IParticle}

\begin{thebibliography}{10}
	
	\bibitem{mostafa1987modeling}
	AA~Mostafa and HC~Mongia.
	\newblock On the modeling of turbulent evaporating sprays: Eulerian versus
	lagrangian approach.
	\newblock {\em International Journal of Heat and Mass Transfer},
	30(12):2583--2593, 1987.
	
	
	\bibitem{faeth1996spray}
	GM~Faeth.
	\newblock Spray combustion phenomena. In: Symposium (international) on combustion.
	\newblock {\em Elsevier}, 1996. p. 1593--1612.

	\bibitem{patrick2012modeling}
	Patrick Jenny, Dirk Roekaerts and Nijso  Beishuizen.
	\newblock Modeling of turbulent dilute spray combustion.
	\newblock {\em Progress in Energy and Combustion Science}, 38(6):846--887, 2012.
	
	
	\bibitem{kovetz1969effect}
	A~Kovetz and B~Olund.
	\newblock The effect of coalescence and condensation on rain formation in a
	cloud of finite vertical extent.
	\newblock {\em Journal of the Atmospheric Sciences}, 26(5):1060--1065, 1969.
	
	\bibitem{langer1990new}
	Robert Langer.
	\newblock New methods of drug delivery.
	\newblock {\em Science}, 249(4976):1527--1533, 1990.
	
	\bibitem{boekerdispersion}
	Egbert Boeker and Rienk Van~Grondelle.
	\newblock Dispersion of pollutants.
	\newblock {\em Environmental Physics: Sustainable Energy and Climate Change}, John Wiley $\&$ Sons, New Jersey 2011.
	
	\bibitem{toschi2009lagrangian}
	Federico Toschi and Eberhard Bodenschatz.
	\newblock Lagrangian properties of particles in turbulence.
	\newblock {\em Annual Review of Fluid Mechanics}, 41:375--404, 2009.
	
	\bibitem{adrian2011particle}
	Ronald~J Adrian and Jerry Westerweel.
	\newblock {\em Particle image velocimetry}.
	\newblock Number~30. Cambridge University Press, 2011.
	
	\bibitem{shew2007instrumented}
	Woodrow~L Shew, Yoann Gasteuil, Mathieu Gibert, Pascal Metz, and
	Jean-Fran{\c{c}}ois Pinton.
	\newblock Instrumented tracer for lagrangian measurements in
	rayleigh-b{\'e}nard convection.
	\newblock {\em Review of scientific instruments}, 78(6):065105, 2007.
	

	\bibitem{ma2016reversed}
	Xing Ma, Seungwook Jang, Mihail~N Popescu, William~E Uspal, Albert Miguel-López, Kersten Hahn, Dong-Pyo Kim, and Samuel Sánchez.
	\newblock Reversed janus micro/nanomotors with internal chemical engine.
	\newblock {\em ACS nano}, 10(9):8751--8759, 2016.
	
	\bibitem{baraban2012catalytic}
	 Larysa	Baraban, Denys Makarov, Robert Streubel, Ingolf Monch, Daniel Grimm, Samuel Sanchez, Oliver~G Schmidt.
	\newblock Catalytic Janus motors on microfluidic chip: deterministic motion for targeted cargo delivery.
	\newblock {\em ACS nano}, 6(4):3383--3389, 2012.

	\bibitem{solovev2011light}
	A~A Solovev, E~J Smith, C~C~B Bufon, S Sanchez, O~G Schmidt.
	\newblock Light‐Controlled Propulsion of Catalytic Microengine.
	\newblock {\em Angew. Chem.}, 123:11067–11070, 2011.
	
	\bibitem{walker2006micro}
	Walker, Daniel.
	\newblock Micro autonomous underwater vehicle concept for distributed data collection.
	\newblock Proceedings of {\em OCEANS 2006}, 1--4, 2006.	
	
	\bibitem{wynn2014autonomous}
	Wynn, Russell B and Huvenne, Veerle AI and Le Bas, Timothy P and Murton, Bramley J and Connelly, Douglas P and Bett, Brian J and Ruhl, Henry A and Morris, Kirsty J and Peakall, Jeffrey and Parsons, Daniel R.
	\newblock Autonomous Underwater Vehicles (AUVs): Their past, present and future contributions to the advancement of marine geoscience.
	\newblock {\em Marine Geology}, 352:451--468, 2014.
	
	\bibitem{calzavarini2018propelled}
	Calzavarini, Enrico and Huang, Yongxiang X and Schmitt, Fran{\c{c}}ois G and Wang, Lipo P.
	\newblock Propelled micro-probes in turbulence.
	\newblock {\em arXiv preprint arXiv:1802.00189}, 2018.
		
	\bibitem{basso2017disposable}
	T~C Basso, M Iovieno, S Bertoldo, G Perotto, A Athanassiou, F Canavero, G Perona, and D Tordella.
	\newblock Disposable radiosondes for tracking Lagrangian fluctuations inside warm clouds.
	\newblock {\em Antennas and Propagation in Wireless Communications (APWC), 2017 IEEE-APS Topical Conference on}, 189--192, 2017.	

	\bibitem{argo}
	D. Roemmich, G.C. Johnson, S. Riser, R. Davis, J. Gilson, W.B. Owens, S.L. Garzoli, C. Schmid, and M. Ignaszewski.
	\newblock The Argo Program: Observing the global ocean with profiling floats.
	\newblock {\em Oceanography}, 22:34--43, 2009.	
	
	\bibitem{weather}
	I. Durre, R. S. Vose, and D. B. Wuertz.
	\newblock Overview of the Integrated Global Radiosonde Archive.
	\newblock {\em Journal of Climate}, 19:53--68, 2006.	

	\bibitem{muinos2018reinforcement}
	Mui{\~n}os-Landin, Santiago and Ghazi-Zahedi, Keyan and Cichos, Frank.
	\newblock Reinforcement Learning of Artificial Microswimmers.
	\newblock {\em arXiv preprint arXiv:1803.06425}, 2018.
	

    \bibitem{sarthou2005diatoms}
	G\'eraldine Sarthou, Klaas R. Timmermans, St\'ephane Blain, and Paul Tr\'eguer.
	\newblock Growth physiology and fate of diatoms in the ocean: a review.
	\newblock {\em Journal of Sea Research}, 53:25--42, 2005.

    \bibitem{gemmell2016buoyancy}
	Brad J. Gemmell, Genesok Oh, Edward J. Buskey, and Tracy A. Villareal.
	\newblock Dynamic sinking behaviour in marine phytoplankton: rapid changes in buoyancy may aid in nutrient uptake.
	\newblock {\em Proc. R. Soc. B}, 283:20161126, 2016.

	
	\bibitem{walsby1994gas}
	A~E Walsby.
	\newblock Gas vesicles.
	\newblock {\em Microbiological reviews}, 58(1):94--144, 1994.

	\bibitem{arrieta2015microscale}
		J Arrieta, A Barreira, I Tuval.
		\newblock Microscale patches of nonmotile phytoplankton.
		\newblock {\em Physical review letters}, 114(12):128102, 2015.
		
	
	\bibitem{maxey1983equation}
	Martin~R Maxey and James~J Riley.
	\newblock Equation of motion for a small rigid sphere in a nonuniform flow.
	\newblock {\em The Physics of Fluids}, 26(4):883--889, 1983.
	
	\bibitem{bec2005multifractal}
	J{\'e}r{\'e}mie Bec.
	\newblock Multifractal concentrations of inertial particles in smooth random
	flows.
	\newblock {\em Journal of Fluid Mechanics}, 528:255--277, 2005.
	
	\bibitem{balkovsky2001intermittent}
	E~Balkovsky, Gregory Falkovich, and A~Fouxon.
	\newblock Intermittent distribution of inertial particles in turbulent flows.
	\newblock {\em Physical Review Letters}, 86(13):2790, 2001.
	
	\bibitem{bec2010turbulent}
	J~Bec, L~Biferale, AS~Lanotte, Andrea Scagliarini, and F~Toschi.
	\newblock Turbulent pair dispersion of inertial particles.
	\newblock {\em Journal of Fluid Mechanics}, 645:497--528, 2010.
	
	\bibitem{biferale2005particle}
	Luca Biferale, Guido Boffetta, Antonio Celani, Alessandra Lanotte, and Federico
	Toschi.
	\newblock Particle trapping in three-dimensional fully developed turbulence.
	\newblock {\em Physics of Fluids}, 17(2):021701, 2005.
	
	\bibitem{douady1991direct}
	S~Douady, Y~Couder, and ME~Brachet.
	\newblock Direct observation of the intermittency of intense vorticity
	filaments in turbulence.
	\newblock {\em Physical Review Letters}, 67(8):983, 1991.
	
	\bibitem{squires1991preferential}
	Kyle~D Squires and John~K Eaton.
	\newblock Preferential concentration of particles by turbulence.
	\newblock {\em Physics of Fluids A: Fluid Dynamics}, 3(5):1169--1178, 1991.
	
	\bibitem{eaton1994preferential}
	John~K Eaton and JR~Fessler.
	\newblock Preferential concentration of particles by turbulence.
	\newblock {\em International Journal of Multiphase Flow}, 20:169--209, 1994.
	
	\bibitem{monchaux2012analyzing}
	Romain Monchaux, Mickael Bourgoin, and Alain Cartellier.
	\newblock Analyzing preferential concentration and clustering of inertial
	particles in turbulence.
	\newblock {\em International Journal of Multiphase Flow}, 40:1--18, 2012.
	
	\bibitem{bec2007heavy}
	Jeremie Bec, Luca Biferale, Massimo Cencini, Alessandra Lanotte, Stefano
	Musacchio, and Federico Toschi.
	\newblock Heavy particle concentration in turbulence at dissipative and
	inertial scales.
	\newblock {\em Physical review letters}, 98(8):084502, 2007.
	
	\bibitem{boffetta2004large}
	G~Boffetta, F~De~Lillo, and A~Gamba.
	\newblock Large scale inhomogeneity of inertial particles in turbulent flows.
	\newblock {\em Physics of Fluids}, 16(4):L20--L23, 2004.
	
	\bibitem{calzavarini2006microbubble}
	L~Biferale, A~Scagliarini and F~Toschi. 
	\newblock On the measurement of vortex filament lifetime statistics in turbulence.
	\newblock  {\em Physics of Fluids}, 22.6: 065101 (2010).
	
	\bibitem{milenkovic2007bubble}
	Rade~{\v{Z}} Milenkovi{\'c}, Beat Sigg, and George Yadigaroglu.
	\newblock Bubble clustering and trapping in large vortices. part 2:
	Time-dependent trapping conditions.
	\newblock {\em International Journal of Multiphase Flow}, 33(10):1111--1125,
	2007.
		
	\bibitem{colabrese2017flow}
	Simona Colabrese, Kristian Gustavsson, Antonio Celani, and Luca Biferale.
	\newblock Flow navigation by smart microswimmers via reinforcement learning.
	\newblock {\em Phys. Rev. Lett.}, 118:158004, 2017.
	
	\bibitem{gustavsson2017finding}
	Kristian Gustavsson, Luca Biferale, Antonio Celani, and Simona Colabrese.
	\newblock Finding Efficient Swimming Strategies in a Three Dimensional Chaotic
	Flow by Reinforcement Learning.
	\newblock {\em Eur. Phys. J. E}, 40:110, 2017.
	
	\bibitem{gazzola2016learning}
	Mattia Gazzola, Andrew~A Tchieu, Dmitry Alexeev, Alexia de~Brauer, and Petros
	Koumoutsakos.
	\newblock Learning to school in the presence of hydrodynamic interactions.
	\newblock {\em Journal of Fluid Mechanics}, 789:726--749, 2016.
	
	\bibitem{gazzola2014reinforcement}
	Mattia Gazzola, Babak Hejazialhosseini, and Petros Koumoutsakos.
	\newblock Reinforcement learning and wavelet adapted vortex methods for simulations of self-propelled swimmers.
	\newblock {\em SIAM Journal on Scientific Computing}, 36.3:B622--B639, 2014.
		
	\bibitem{novati2017synchronisation}
	Guido Novati, Siddhartha Verma, Dmitry Alexeev, Diego Rossinelli, Wim~M van~Rees,  and Petros Koumoutsakos.
	\newblock Synchronisation through learning for two self-propelled swimmers.
	\newblock {\em Bioinspiration \& biomimetics}, 12.3:036001, 2017.
		
	\bibitem{verma2018efficient}
	Siddhartha Verma, Guido Novati, and Petros Koumoutsakos.
	\newblock Efficient collective swimming by harnessing vortices through deep reinforcement learning.
	\newblock {\em arXiv preprint}, arXiv:1802.02674, 2018.

		
	\bibitem{reddy2016learning}
	Gautam Reddy, Antonio Celani, Terrence~J Sejnowski, and Massimo Vergassola.
	\newblock Learning to soar in turbulent environments.
	\newblock {\em Proceedings of the National Academy of Sciences}, page
	201606075, 2016.
	
	\bibitem{sutton2017reinforcement}
	R.S. Sutton and A.G. Barto.
	\newblock {\em Reinforcement Learning: An Introduction}.
	\newblock MIT Press, Cambridge 2017.
	
	\bibitem{tesauro1995temporal}
	Gerald Tesauro.
	\newblock Temporal difference learning and TD-Gammon.
	\newblock {\em Communications of the ACM}, 38(3):58--68, 1995.

	\bibitem{kaelbling1996reinforcement}
	Leslie Kaelbling, Littman Pack, L Michael, and  Andrew~W Moore.
	\newblock Reinforcement learning: A survey.
	\newblock {\em Journal of artificial intelligence research}, 4:237--285, 1996. 

	\bibitem{auton1988force}
	TR~Auton, JCR Hunt, and M~Prud'Homme.
	\newblock The force exerted on a body in inviscid unsteady non-uniform
	rotational flow.
	\newblock {\em Journal of Fluid Mechanics}, 197:241--257, 1988.
	
	\bibitem{babiano2000dynamics}
	Armando Babiano, Julyan~HE Cartwright, Oreste Piro, and Antonello Provenzale.
	\newblock Dynamics of a small neutrally buoyant sphere in a fluid and targeting
	in hamiltonian systems.
	\newblock {\em Physical Review Letters}, 84(25):5764, 2000.
	
	\bibitem{gatignol1983faxen}
	Ren{\'e}e Gatignol.
	\newblock The fax{\'e}n formulas for a rigid particle in an unsteady
	non-uniform stokes-flow.
	\newblock {\em Journal de M{\'e}canique th{\'e}orique et appliqu{\'e}e},
	2(2):143--160, 1983.
	
	\bibitem{watkins1992q}
	Christopher~JCH Watkins, and Peter Dayan.
	\newblock Q-learning.
	\newblock {\em Machine learning}, 8(3-4):279--292, 1992.
	
	
	\bibitem{dombre1986chaotic}
	Thierry Dombre, Uriel Frisch, John~M Greene, Michel H{\'e}non, A~Mehr, and
	Andrew~M Soward.
	\newblock Chaotic streamlines in the abc flows.
	\newblock {\em Journal of Fluid Mechanics}, 167:353--391, 1986.
	
\end{thebibliography}

\end{document}